\documentclass[lettersize,journal]{IEEEtran}
\usepackage{amsmath,amsfonts}
\usepackage{algorithmic}
\usepackage[ruled,vlined]{algorithm2e} 
\usepackage{array}
\usepackage[caption=false,font=normalsize,labelfont=sf,textfont=sf]{subfig}
\usepackage{textcomp}
\usepackage{stfloats}
\usepackage{url}
\usepackage{verbatim}
\usepackage{graphicx}
\usepackage{xcolor}
\usepackage{booktabs}
\usepackage{multirow} 
\usepackage[table,xcdraw]{xcolor}
\usepackage[marginal]{footmisc}
\def\BibTeX{{\rm B\kern-.05em{\sc i\kern-.025em b}\kern-.08em
    T\kern-.1667em\lower.7ex\hbox{E}\kern-.125emX}}

\usepackage{balance}

\newcommand{\sysname}{Janus\xspace}

\begin{document}

\title{Enabling Agile Ambient IoT Networking \\ via a Parameterized Hybrid Radio}

\author{Jiazhen Lei, Fengyuan Zhu, Tianze Cao, Yuxin Sha, Linling Zhong, Wenhui Li, Bingbing Wang, Zeming Yang, Jinyang Sun, Yibin Deng,~\IEEEmembership{Student Member,~IEEE}, Xiaohua Tian,~\IEEEmembership{Senior Member,~IEEE}

\thanks{\hspace{3mm}The work was supported in part by the National Natural Science Foundation of China under Grant 92567202. \textit{(Corresponding author: Xiaohua Tian.)}}

\thanks{\hspace{3mm}Jiazhen Lei, Fengyuan Zhu, Tianze Cao, Yuxin Sha, Linling Zhong, Wenhui Li, Zeming Yang, Jinyang Sun, Yibin Deng and Xiaohua~Tian are with the School of Electronic Information and Electrical Engineering, Shanghai Jiao Tong University, Shanghai 200240, China. Bingbing Wang is with the School of Computer Science and Artificial Intelligence,  Zhengzhou University, Zhengzhou 450001, China. E-mail: \{leijz, jsqdzhufengyuan, ctzzsjtu, cindysha, uranium\_zll, lwh1999, yzm\_sjtu, jysun, carlos.dengyibin, xtian\}@sjtu.edu.cn, wangbingbing@zzu.edu.cn.}
}

\markboth{IEEE/ACM TRANSACTIONS ON NETWORKING}%
{Zhu \MakeLowercase{\textit{et al.}}: Bare Demo of IEEEtran.cls for Computer Society Journals}

\maketitle
\begin{abstract}
The emergence of Ambient IoT signals a paradigm shift toward massive batteryless networking. However, the absence of an agile physical layer substrate remains a fundamental barrier to research and standardization. Current testbeds are hindered by decoupled radio paths, high static power, and cumbersome control methods, which stifle rapid protocol prototyping. In this paper, we present \sysname, the first hybrid active-passive configurable radio architected for agile Ambient IoT networking. \sysname introduces a parameterized architecture that unifies passive and active transmission into a single RF front end, abstracting complex physical layer behaviors into concise parameters. This design enables a system-level control plane for dynamic mode transitions and an energy management plane for fine-grained harvesting across multiple sources. We implement a compact PCB prototype and evaluate its performance across diverse protocol landscapes, including 3GPP A-IoT, IEEE 802.11 AMP, and Bluetooth SIG. Our experimental results demonstrate that \sysname achieves communication performance on par with dedicated radios while significantly reducing configuration overhead. Ultimately, \sysname serves as a versatile enabler for validating emerging protocols and accelerating the standardization of next-generation low-power networks.
\end{abstract}


\section{Introduction}
The Internet of Things is undergoing a fundamental paradigm shift towards Ambient IoT (A-IoT). Envisioned as the bedrock of the 6G era, A-IoT devices are designed to be ubiquitous, maintenance-free, and powered exclusively by energy harvested from the environment (e.g., solar, RF, thermal)~\cite{butt2024ambient}. Recognizing this potential, major standardization bodies, including 3GPP (Rel-19 Ambient IoT~\cite{3gpp_tr38769}), IEEE 802.11 (AMP~\cite{80211AMP}), and the Bluetooth SIG~\cite{bluetoothSIG}, are aggressively defining new specifications to accommodate these ultra-low-power endpoints.

\begin{figure}[t]
    \centering
    \includegraphics[width=0.99\linewidth]{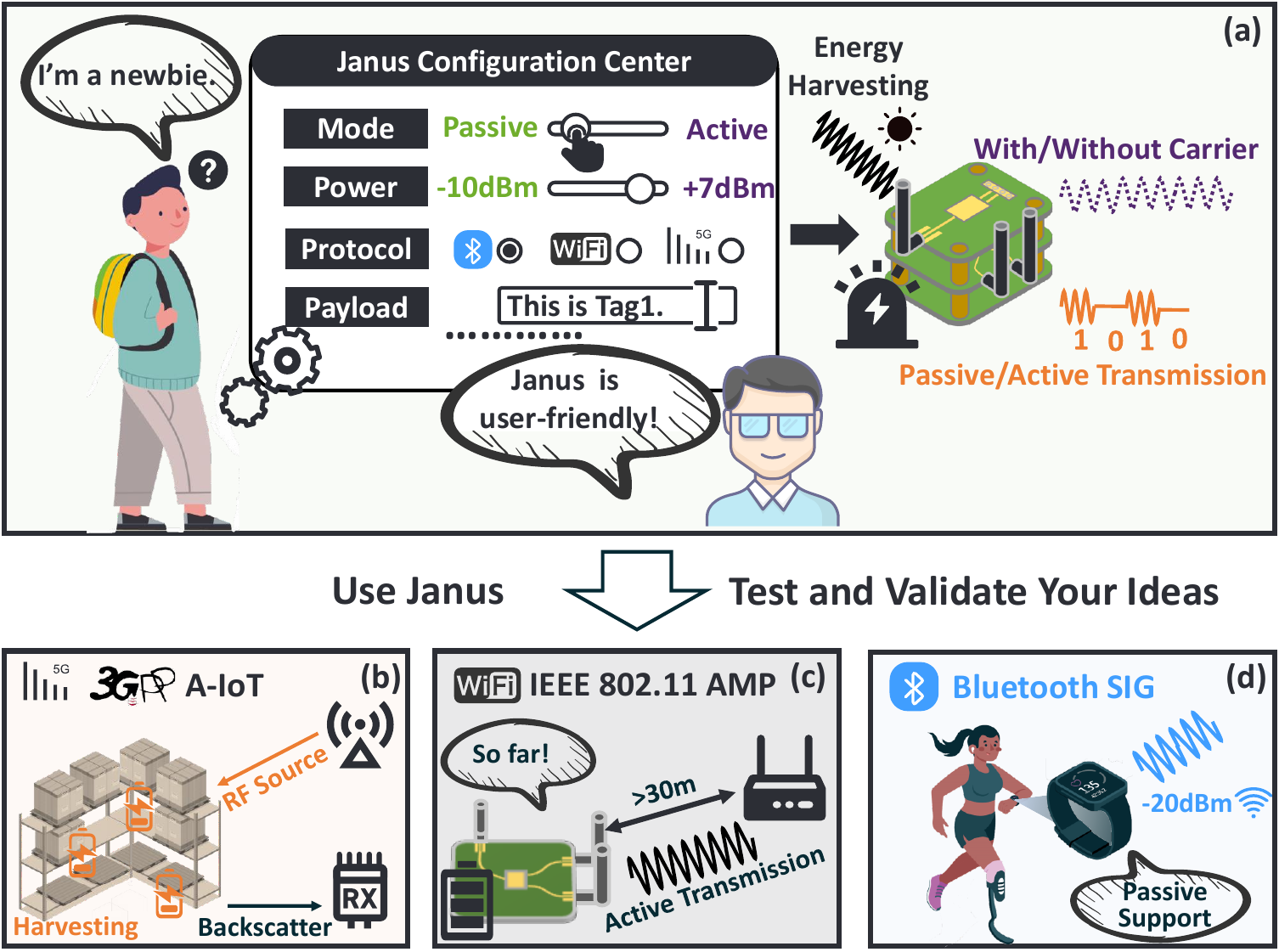}
    \caption{\sysname employs a fully parameterized configuration approach to enable hybrid active-passive communication. It facilitates comprehensive case studies across major standardization organizations, such as 3GPP A-IoT(a), IEEE 802.11 AMP(b), Bluetooth SIG(c).}
    \label{fig:teaser}
\end{figure} 

The successful deployment of A-IoT networking, however, hinges on resolving the inherent trade-off between communication range and power budget. Backscatter communication~\footnote{Backscatter communication represents a passive communication technique. Throughout this paper, we employ the term ``passive communication'' to denote backscatter technology. } has long been the technological cornerstone of battery-free devices due to its microwatt-level power consumption. Yet, its reliance on external RF carriers inherently limits range and reliability due to double-link attenuation~\cite{kellogg2016passive, ambientbs, OFDMAbackscatter, lorabs, interscatter, IBLE, CAB, LScatter, BiScatter, FaB}. Conversely, active transmission offers robust coverage and independence from ambient signals but traditionally demands power budgets incompatible with meager harvested energy~\cite{3gpp_tr38769, 80211AMP}. 
Consequently, a consensus in the research community is that next-generation A-IoT networks are hybrid, encompassing passive, active, and hybrid device classes to accommodate varying energy availability and channel conditions in complex environments.

Despite the urgent demand for hybrid A-IoT protocols, networking research is currently stifled by the lack of a capable hardware substrate. The community requires an \textit{\textbf{efficient, general-purpose, and user-friendly}} testing platform capable of prototyping network systems across multiple communication modes and protocols. Such a platform must support seamless transitions between active and passive communication to evaluate adaptive networking strategies while remaining accessible to researchers with varying hardware expertise. Furthermore, to align with mainstream standards, the system should accommodate sub-$6GHz$ wideband frequencies and integrate native multi-source energy harvesting capabilities.

\begin{table*}[t]
    \centering
    \caption{Comparison Between Different Radio Platforms.}
    \includegraphics[width=0.98\linewidth]{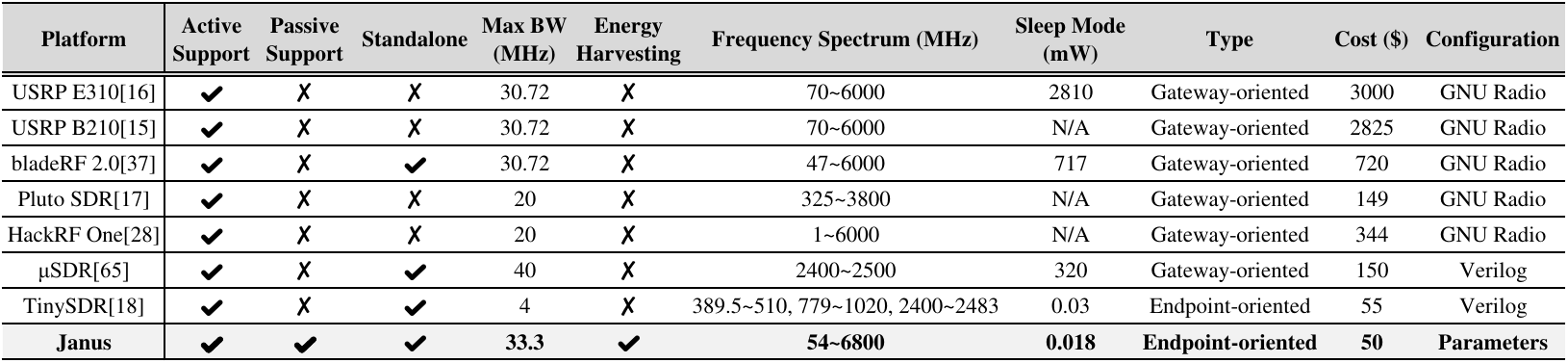}
    \label{table:comparsion_sdr}
\end{table*}

Unfortunately, existing radio platforms fail to bridge this gap. Commercial radios (e.g., USRP~\cite{USRPB210, USRPE310}, Pluto SDR~\cite{plutoSDR}) are primarily designed for gateways; their watt-level power consumption and tethered form factors make them unsuitable for emulating energy-constrained endpoints.
On the other hand, the existing endpoint-oriented platform TinySDR represents a significant step forward by targeting IoT scenarios~\cite{tinysdr}. However, the emerging A-IoT introduces distinctive requirements, specifically hybrid active and passive communication as well as energy harvesting, that extend beyond the design scope of prior platforms oriented towards endpoints.
More critically, developing on these FPGA-based platforms entails a steep learning curve, compelling networking researchers to master Hardware Description Languages (HDL) to modify PHY layer behaviors. This hardware-software coupling significantly slows the iteration of upper-layer protocols. As summarized in Table~\ref{table:comparsion_sdr}, a unified platform that simultaneously addresses agility, energy constraints, and usability is absent.

To dismantle these barriers, we present \sysname, the first hybrid active-passive configurable radio platform specifically architected for the A-IoT networking ecosystem. Unlike traditional gateways or rigid sensor tags, \sysname is a versatile endpoint radio that supports standalone operation in both active and passive modes across sub-$6GHz$ frequencies. It features an intuitive parameter-based interface that eliminates complex HDL programming, democratizing access to physical-layer agility for network protocol designers.

Designing such a platform required overcoming fundamental challenges across circuit architecture, control abstraction, and energy management:

\textbf{1) Unified Active-Passive Hardware Architecture.}
A-IoT networking requires fluid switching between modes to meet performance requirements~\cite{3GPP1}. The fundamental architectural distinction between these modes lies in carrier generation: while backscatter modulates ambient carriers via a simple RF switch, active communication typically relies on power-intensive superheterodyne architectures that require DACs, oscillators, and mixers~\cite{CC2650, CC3350, AD9361, AD9363, MAX2828, AT86RF215}. \sysname unifies these paradigms by employing a novel shared-baseband architecture. It leverages a wideband frequency synthesizer as a local carrier source and reconfigures the RF front-end to drive subsequent transmission, effectively merging active and passive paths into a single, compact substrate.

\textbf{2) User-Friendly Control Plane.}
Rapid protocol prototyping is fundamental to agile networking research. Although toolkits such as GNU Radio~\cite{GNURadio} effectively support gateway radios, adapting endpoint-specific FPGA designs to diverse standards typically requires modifying the underlying hardware logic. This requirement presents a high technical barrier for many researchers. To address this challenge, \sysname abstracts complex physical-layer behaviors into a \textit{Parameterized Control Plane}. By exposing a generic digital baseband, \sysname enables users to configure hybrid communication modes and protocols via a concise set of hardware-independent parameters, thereby eliminating the need for extensive low-level code modification.

\textbf{3) Scalable Energy Management Plane.}
To serve as a faithful testbed for A-IoT, the platform must operate under the constraints of environmental energy harvesting. \sysname incorporates an energy management plane that orchestrates multi-source harvesting (e.g., RF and light) and fine-grained power allocation. This design allows the device to dynamically transition between duty-cycled operation and exhaustion modes based on real-time energy availability, enabling the evaluation of energy-aware networking protocols.

We implement \sysname as a compact PCB prototype and validate its effectiveness through case studies covering protocols from three representative standardization bodies: 3GPP A-IoT, IEEE 802.11 AMP, and Bluetooth SIG. Our experiments demonstrate that \sysname achieves performance parity with dedicated commercial hardware in both modes. In terms of usability, \sysname achieves a $14.6\times$ to $45.6\times$ speedup in configuration efficiency compared to traditional Python-UHD and Verilog-based workflows. Furthermore, relative to state-of-the-art endpoint-oriented radios, \sysname delivers an $8.3\times$ increase in bandwidth and a $1.7\times$ reduction in sleep power consumption, offering the research community a versatile foundation for the next generation of ambient networking.

To summarize, we make the following contributions:
\begin{itemize}
    \item We propose \sysname, the first configurable radio platform tailored for A-IoT endpoints. It uniquely integrates multi-source energy harvesting with a hybrid RF architecture capable of both active and passive communication.
    \item We introduce a series of architectural innovations to enable parametric configuration, including a unified active-passive RF front-end, a customized interaction channel, and a parameterized control plane. This design bridges the gap between hardware efficiency and software flexibility, enabling full system control via intuitive parameters.
    \item We implement the hybrid active-passive radio using a PCB prototype. Our evaluation confirms that \sysname effectively serves as an agile testbed for the development and standardization of A-IoT networking protocols.
\end{itemize}

\textbf{Open-source plan.} 
\sysname is designed for the A-IoT research community and standardization. 
We intend to open-source hardware design and software stack. This release will include schematics/layouts, firmware, host-side software, comprehensive user manuals, and example scripts.
All data and code will be publicly available at: \url{https://github.com/Jiazhen-Lei/Janus}

\section{Primer \& Rethinking }
\subsection{Primer on Software-defined Baseband}
In passive communication networks, our prior work SD-PHY \cite{sdphy} parametrically models backscattered signals with arbitrary IoT standards. Specifically, for a desired backscattered signal conforming to an arbitrary IoT protocol, the reflection coefficient sequence can always be derived based on the excitation signal. The backscatter sequence is generated by selecting and switching between different impedances through the baseband signal. Therefore, fine-grained manipulation of baseband signal generation suffices to produce arbitrary backscatter waveforms. 
\begin{equation}\label{eq:2}
    \phi(n)=2\pi\cdot(\frac{1}{2}a_2n^2+a_1n+a_0)
\end{equation}
\begin{figure}[t]
    \centering
    \includegraphics[width=0.98\linewidth]{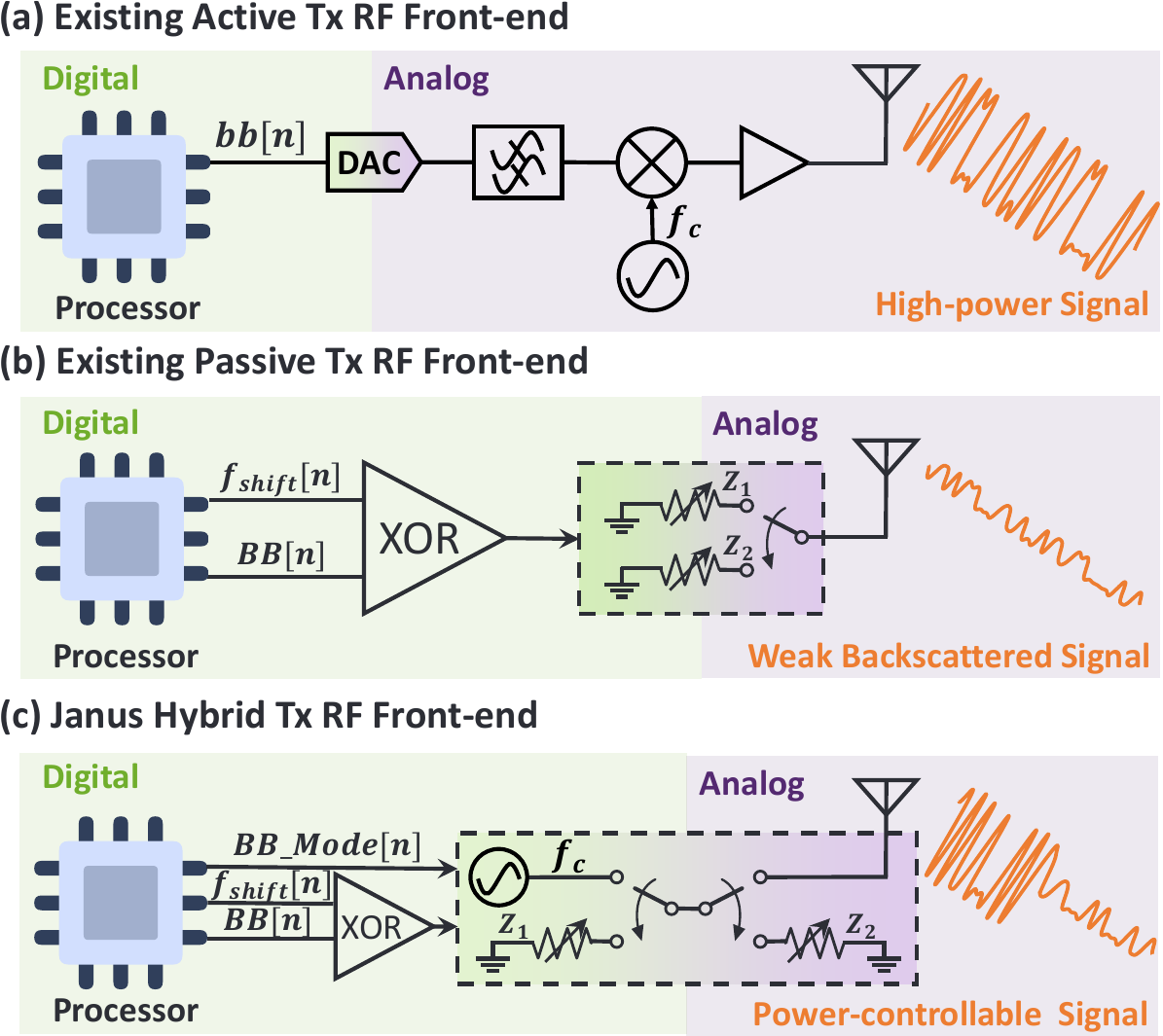}
    \caption{Comparison between existing RF Tx front-ends and \sysname.}
    \label{fig:rf_frontend}
\end{figure} 
\begin{figure}[t]
    \centering
    \includegraphics[width=0.99\linewidth]{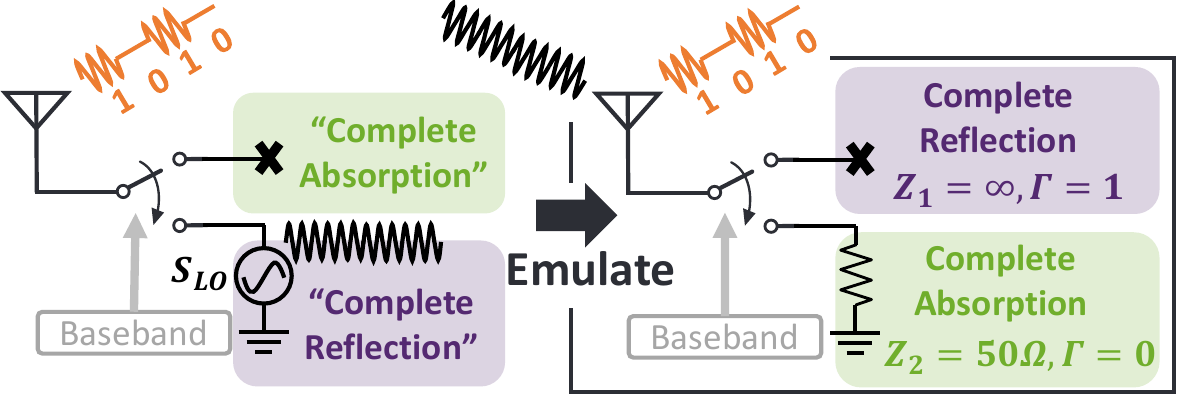}
    \caption{By toggling the RF switch, the alternation of the local carrier path can emulate the variation in reflection coefficient induced by impedance switching.}
    \label{fig:carrier_pattern}
\end{figure} 
In practical deployments, tags struggle to achieve continuous reflection coefficients and typically perform a quantization process. The baseband signal serves as the control signal for RF switches to connect different impedances. Eq.~\ref{eq:1} can be used to describe the process of quantizing and switching reflection coefficients via the baseband signal.
\begin{equation}\label{eq:1}
    \Gamma(n)=\textbf{Q}[Re(S_{BB}(n))]=\textbf{Q}[cos(\phi(n))] 
\end{equation}

The baseband signal $S_{BB}(n)$, after extracting its real component, retains only the phase term $cos(\phi(n))$. The operator $\textbf{Q}[\cdot]$ encapsulates the quantization and switching logic that maps the baseband signal to RF switch control. The discrete-time index $n$ ranges from $0$ to $N_s-1$, where $N_s$ denotes the number of samples per symbol. The phase term $cos(\phi(n))$ can be expressed with a second-order Taylor series expansion as follows:

We only need four parameters, $a_2$, $a_1$, $a_0$ and $N_s$, to describe symbols containing a large number of sampling points to implement PSK, FSK and CCS modulation. 1) To realize PSK, we set parameter $a_2=0$ and $a_1=0$, then we embed the payload on $a_0$, e.g., $a_0=\frac{0}{2\pi}$ and $a_0=\frac{\pi}{2\pi}$. 2) To realize FSK, we set $a_2$ = 0 and leave $a_0$ fixed, then we embed the payload on $a_1$. 3) To realize CSS, we set $a_2$ to be the chirp rate, $a_1$ to be the starting frequency, and $a_0$ fixed. Parameter $N_s$ can be configured according to the symbol rate.
In fact, SD-PHY deploys a software-defined baseband (SD-Baseband) that parameterizes the synchronization module, modulation module, and frequency synthesizer. However, this is not enough to constitute a complete software-defined radio platform.
\begin{figure*}[t]
    \centering
    \includegraphics[width=0.98\linewidth]{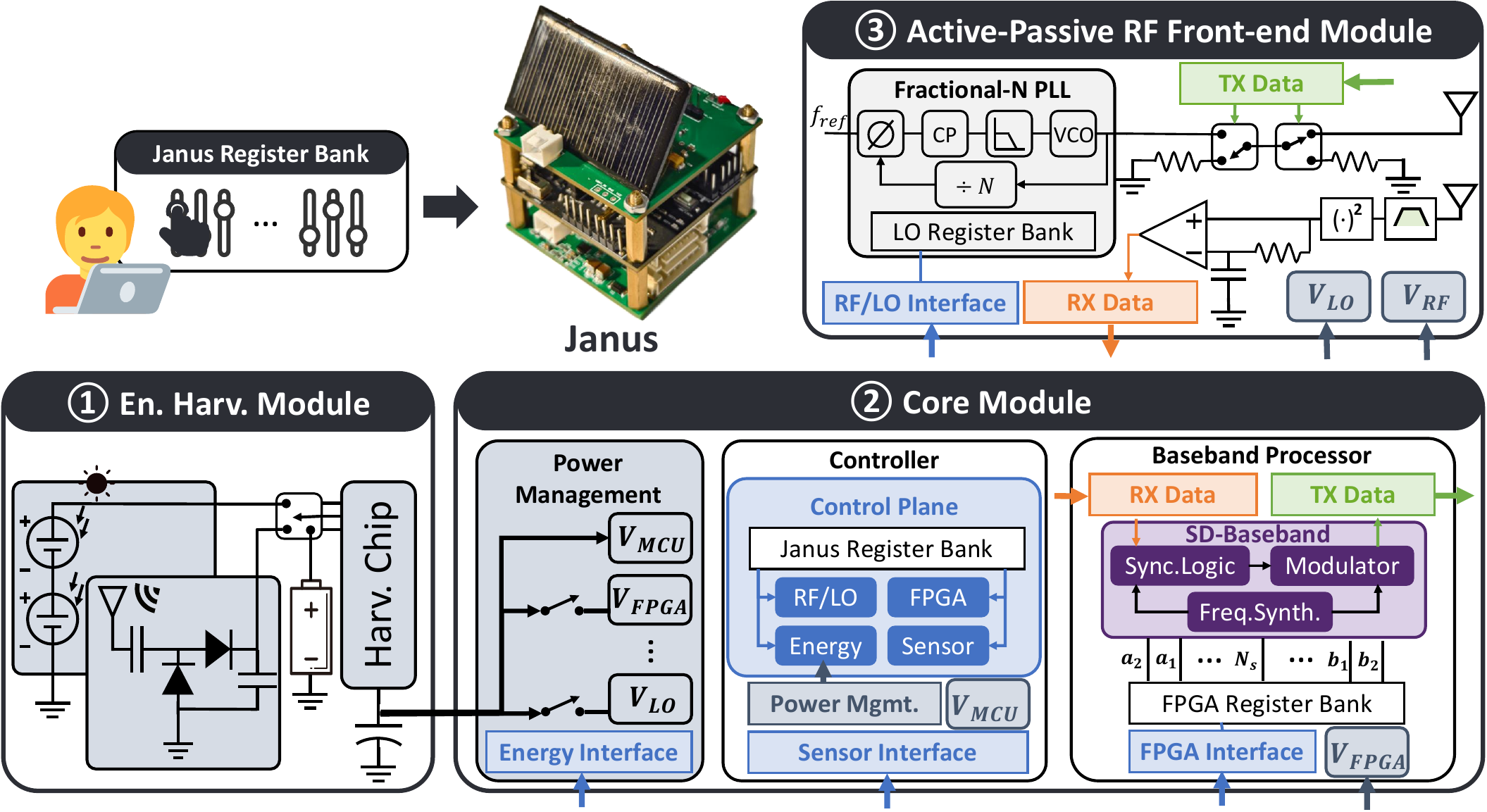}
    \caption{\sysname System Block Diagram. A complete system diagram showing all of the components of \sysname. The architecture primarily contains three modules: 1) a multi-source energy harvesting module with extensible interface; 2) a core module with energy management, control, and baseband processing capabilities; 3) an active-passive RF front-end with switchable operation modes.}
    \label{fig:overview}
\end{figure*} 

\subsection{Standalone RF front-end}
An A-IoT network employs active communication for IoT devices in harsh environments. However, as Fig.~\ref{fig:rf_frontend} (a) and (b) show, active and passive schemes differ fundamentally in the RF front-end design.
Active communication requires power-intensive components like DACs, frequency synthesizers, and mixers for upconversion. In contrast, passive backscatter systems simplify the RF front-end by directly controlling an RF switch for modulation via impedance toggling.
Consequently, the baseband bitstream $BB[n]$ in passive communication is generated via $x$-bit quantization before output, which differs fundamentally from the signal generation principles of the traditional active baseband $bb[n]$.
Furthermore, to achieve frequency shifting in backscatter~\cite{kellogg2016passive}, the baseband signal $BB[n]$ typically undergoes an XOR operation with a shifting frequency sequence $f_{shift}[n]$ to control the impedance switching, thereby translating the backscattered signal to the target frequency band.

To seamlessly integrate these modes on commercial radio platforms typically necessitates an auxiliary RF path for backscatter, which inflates hardware complexity. Moreover, conventional platforms impose a significant development burden, requiring host-based tools (e.g., GNU Radio) or distinct Verilog modules for each protocol. Notably, even endpoint-oriented solutions like TinySDR~\cite{tinysdr} remain bound by this rigid requirement for protocol-specific baseband implementations.

\subsection{Rethinking Carrier Generation Patterns}
We identify that both communication paradigms share fundamental baseband processing requirements while differing primarily in carrier generation patterns and quantitative strategies. This observation leads to our question: \textbf{\textit{Can the parameterized baseband signal approach be effectively extended to active communication scenarios?}}
Our investigation and design conclusively demonstrates that this is indeed \textbf{FEASIBLE}.

Consider a 1-bit ($x=1$) backscatter tag with reflection coefficients $\Gamma_1$ and $\Gamma_2$. Reviewing antenna theory, the reflection coefficient can be expressed as
\begin{equation}\label{eq:3}
    \Gamma_i=\frac{Z_i-Z_0}{Z_i+Z_0}
\end{equation}
where $Z_0$ represents antenna impedance (typically $50\Omega$) and $Z_i$ denotes the device impedance. By configuring $Z_1=\infty$ (open circuit) and $Z_2 = Z_0 = 50\Omega$, we obtain $\Gamma_1=1$ and $\Gamma_1=0$. The RF switch toggles between these at $\frac{1}{T_s}$ intervals, effectively multiplying the incident signal with a baseband-controlled square wave.

For active communication, replacing $Z_2$ with local oscillator $S_{LO}$ (LO) while maintaining RF switching achieves the same $\Gamma$ states, as shown in Fig.~\ref{fig:carrier_pattern}. The switch acts as a mixer, producing the desired product term between baseband and oscillator signals. This unified approach enables seamless mode switching between active and passive communication.

To extend this paradigm to active communication, we substitute the absorptive load $Z_2$ with a Local Oscillator source ($S_{LO}$), while retaining the identical RF switching architecture. As illustrated in Fig.~\ref{fig:carrier_pattern}, the switch effectively functions as a mixer, generating the desired product term between the baseband signal and the local oscillator. 

This unified architectural approach offers a profound advantage beyond mere hardware simplification: it creates a consistent abstraction layer for hybrid networking. By generalizing the parameterized baseband to active transmission, we decouple the protocol definition from the underlying physical signal generation mechanisms. This allows upper-layer protocols to seamlessly toggle between transmission modes using a unified set of hardware-agnostic parameters, thereby eliminating the rigid hardware-software coupling that currently hinders the rapid prototyping of adaptive A-IoT systems.

\section{\sysname Architecture Design}
To bridge the gap between rigid commercial testbeds and the agile requirements of A-IoT, we introduce the \sysname architecture. As illustrated in Fig.~\ref{fig:overview}, the system is designed around three fundamental design principles: a unified hybrid RF substrate, a parameterized control abstraction, and an energy-aware management plane. Unlike traditional radios that decouple active and passive paths, \sysname integrates them into a holistic framework, enabling dynamic mode switching while adhering to the stringent power budgets of battery-free computing. The following section details the architectural evolution toward a fully agile, hybrid A-IoT radio.

\subsection{Hybrid Active-Passive Radio}\label{Sec:3.1}
The physical layer of A-IoT requires a versatile RF front-end capable of operating across the sub-6 GHz spectrum in both active and passive modes. The primary design challenge lies in integrating the distinct paradigms of high-power carrier generation and zero-power impedance modulation while avoiding the hardware redundancy and power overhead typically associated with gateway-oriented radios.

%
\subsubsection{Unified RF Front-end}\leavevmode

\textbf{Wideband Carrier Synthesis.}
Existing radio platforms typically employ transceiver chips with integrated frequency synthesizers for LO generation and baseband upconversion. As Table~\ref{table:comparsion_rfsyn} shows, current commercial solutions prove unsuitable for \sysname.
First, mainstream radios~\cite{USRPB210,USRPE310} typically leverage high-performance transceivers such as the AD936x-series\cite{AD9361,AD9363} and MAX2828\cite{MAX2828}. While these components offer robust sub-$6GHz$ support, their power consumption (reaching $1.7W$) is primarily tailored for gateway applications rather than energy-constrained A-IoT endpoints. Second, although TinySDR\cite{tinysdr} effectively reduces power by utilizing the AT86RF215 transceiver\cite{AT86RF215}, its frequency synthesizer is designed for standard Sub-GHz and $2.4GHz$ ISM bands, leaving coverage gaps for cellular A-IoT frequencies such as 3GPP Band 3 ($1805-1880MHz$)\cite{3gpp_tr38769}. Finally, platforms like HackRF One\cite{HackRF} achieve wideband coverage via a two-stage upconversion architecture (MAX2839\cite{MAX2839} and RFFC5072\cite{RFFC5072}), a design choice that incurs a power overhead exceeding the stringent budgets of battery-free devices.

Transceiver chips incorporate more redundant components. Since broadband RF switches can functionally replace DACs and mixers, we optimize the design by focusing solely on carrier generation. For programmable sub-$6GHz$ coverage, we adopt a fractional-N PLL architecture that offers $Hz$-level frequency resolution without compromising clock stability. 
Moreover, the state-of-the-art fractional-N PLL in $65nm$ CMOS consumed $17.9mW$~\cite{jo2023135fs}, covering $600MHz-7.7GHz$.
During the \sysname prototype design stage, we select ADF4355~\cite{ADF4355} for its wide RF output frequency range from $54MHz$ to $6.8GHz$.
\begin{table}[t]
    \centering
    \caption{\centering Comparison Between Different \\ Active RF Front-ends.}
    \includegraphics[width=0.98\linewidth]{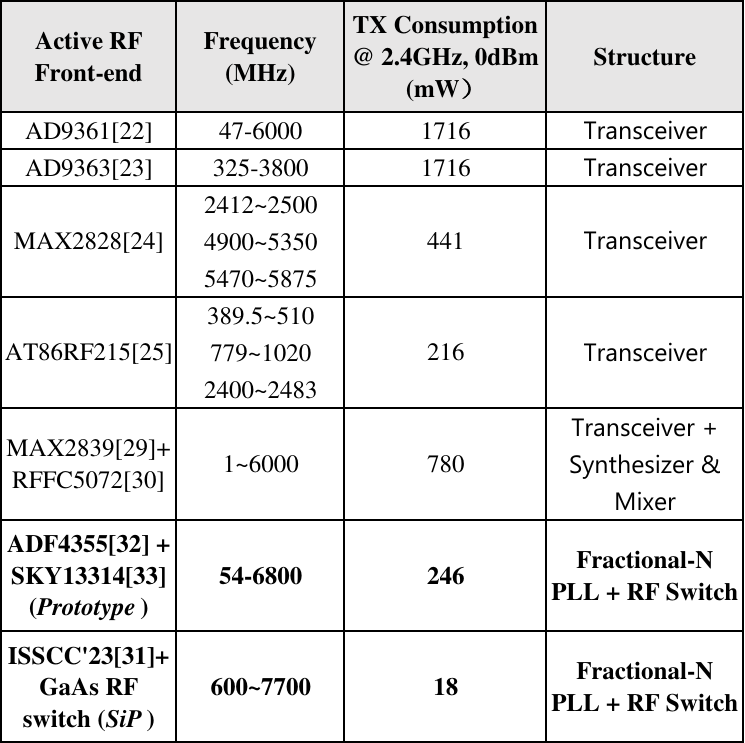}
    \label{table:comparsion_rfsyn}
\end{table}

\textbf{Dual-Mode Modulation Interface.}
To unify the active and passive transmission paths, we exploit the functional duality of the RF switch. In passive mode, the switch modulates impedance to reflect ambient signals; in active mode, we reconfigure it to act as a DAC and a mixer, modulating the local carrier generated by the PLL. This shared-component approach minimizes insertion loss and hardware complexity, effectively creating a mode-agnostic modulation interface.

The RF switch needs to cover sub-$6GHz$, and the switching response on/off time frequency needs to be above $20MHz$ to meet the bandwidth requirement of IEEE 802.11.
\sysname employs a GaAs pHEMT-based SPDT RF switch. The pHEMT technology enables rapid switching and wide spectral coverage for high-dynamic-range applications. In prototype design, we choose SKY13314-373LF~\cite{SKY13314} with sub-$6GHz$ coverage and $33.3MHz$ switching speed.
Furthermore, through system-in-package (SiP) integration of the GaAs RF switch with the PLL, the power consumption of the active RF front-end can be reduced below $18mW$.

\textbf{Low-power OOK receiver.}
As a novel device class supplementing existing network infrastructure, A-IoT devices must prioritize low power consumption without compromising compatibility with legacy systems. Both 3GPP A-IoT~\cite{{3gpp_tr38769}} and 802.11AMP~\cite{80211AMP} widely employ an OFDM-based waveform to synthesize OOK modulation at the gateway for the downlink. In this paradigm, gateways can leverage existing DFT-based OFDM transmitters to generate downlink OOK waveforms, thereby allowing A-IoT devices to utilize energy-efficient non-coherent demodulation techniques for signal reception. Specifically for \sysname, a configurable radio tailored for A-IoT applications, we implement a low-power OOK receiver comprising an envelope detector, an RC integrator, and a comparator.

%
\subsubsection{Resource-Constrained Baseband Processing}\leavevmode

The unified RF front-end architecture necessitates a compute substrate capable of generating heterogeneous baseband signals while operating under a stringent microwatt-level power envelope. Unlike traditional radios that rely on power-hungry FPGAs for complex PHY processing, \sysname decouples high-speed modulation from system-level management through a dual-core heterogeneous controller-processor architecture.

\begin{figure}[t]
    \centering
    \includegraphics[width=0.98\linewidth]{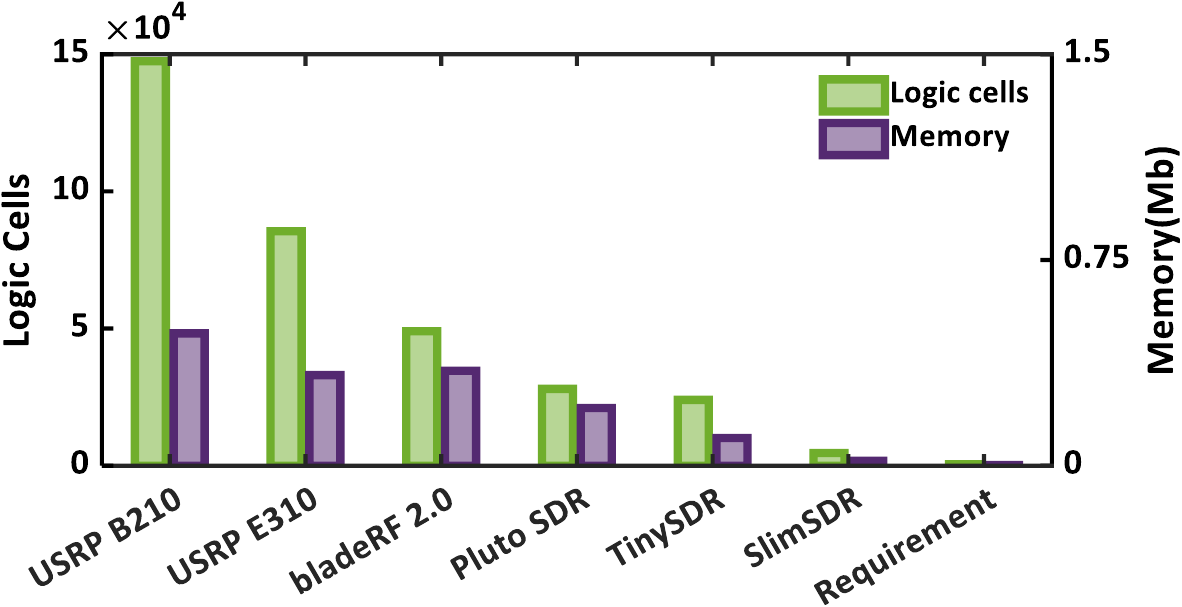}
    \caption{The baseband processor resources of conventional radios far exceed \sysname's requirements.}
    \label{fig:comparsion_fpga}
\end{figure}

\textbf{Lean Baseband Processing.}
To achieve cross-protocol agility without the overhead of bitstream-level reconfiguration, \sysname leverages an SD-Baseband that integrates a synchronized wake-up receiver, a universal modulator, and a digital frequency synthesizer~\cite{sdphy, SDPHYopensource}. By shifting from manual bitstream manipulation to a parameter-driven execution model, we eliminate the need for protocol-specific HDL synthesis. As demonstrated by the post-layout simulations shown in Fig. \ref{fig:comparsion_fpga}, the SD-Baseband occupies only 548 logic cells and 21.6 KB of memory, achieving resource efficiency that is orders of magnitude superior to commercial gateway-oriented radios. During the prototype validation phase, our implementation on a Gowin GW1N FPGA~\cite{gowin} maintains a sub-$5mW$ active power profile, ensuring compatibility with intermittent energy sources.

\textbf{Ultra-low-power System Management.}
To orchestrate the state transitions and parameter interpretation, we incorporate a dedicated low-power MCU as the central management unit. The MCU's selection is dictated by the requirement for flexible I/O capability and non-volatile state retention during deep-sleep cycles. We employ a 16-bit RISC MSP430FR5969~\cite{msp430fr5969} with 64KB of FRAM, which serves as the host for both the control plane (\S\ref{Sec:3.2}) and the energy management plane (\S\ref{Sec:3.3}). This hierarchy enables a zero-programming user interface, where complex hardware behaviors are abstracted into high-level register configurations.

\subsubsection{Integrated Hybrid Transmission Paths}\leavevmode

\begin{figure}[t]
    \centering
    \includegraphics[width=0.98\linewidth]{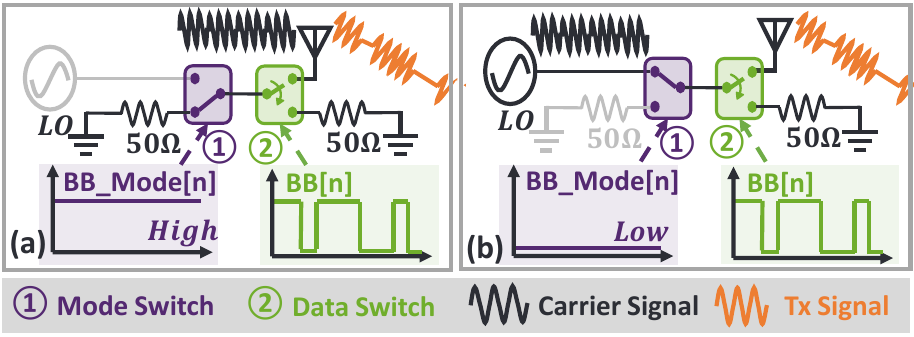}
    \caption{\sysname adjusts \textit{Mode Switch} and \textit{Data Switch} to implement passive (a) and active (b) data paths.}
    \label{fig:data_channel}
\end{figure}

A critical challenge in hybrid radio design is ensuring signal integrity across fundamentally different transmission modes within a minimal hardware footprint. To address this, \sysname unifies the data path through a dual-switch topology (Fig.~\ref{fig:data_channel}) that reuses the same digital baseband interface for both modes.

In passive mode, the mode switch connects to a $50\Omega$ resistor, while the baseband signal toggles the data switch between reflection coefficients $\Gamma=0$ and $\Gamma=1$. In active mode, the mode switch routes to the PLL, and the baseband signal controls the data switch to alternate between the antenna and the $50\Omega$ load. The $50\Omega$ load specifically prevents PLL damage due to open-circuit conditions.

%
\subsection{Parameter-Driven System Orchestration}\label{Sec:3.2}

Commercial radios~\cite{USRPB210,USRPE310,bladeRF,plutoSDR} typically rely on host-based signal processing using open-source toolkits (e.g., GNU Radio~\cite{GNURadio}) or custom code blocks. While flexible, this architecture generally necessitates a wired connection to a host processor, which can constrain standalone deployment in IoT scenarios. Similarly, endpoint-oriented platforms~\cite{tinysdr} often require full FPGA bitstream re-synthesis for even minor protocol adjustments, imposing a steep learning curve for networking researchers.

\sysname resolves this fundamental trade-off by extending the parameterized configuration model, which was previously confined to passive baseband processing, into a comprehensive full-stack radio architecture. By decoupling physical-layer execution from system-level management via a customized MCU-FPGA interaction protocol and a parameterized control plane, the system achieves dynamic runtime agility while eliminating the requirement for hardware-level reprogramming.

%
\subsubsection{Interaction Channel}\leavevmode

The architectural foundation of agility is an ``MCU+FPGA'' heterogeneous substrate, where the FPGA serves as a high-speed baseband engine and the MCU acts as a programmable orchestrator. To bridge these two domains, we design a custom interaction channel that replaces the static firmware-loading approach used in prior works~\cite{tinysdr}.

\begin{figure}[t]
    \centering
    \includegraphics[width=0.98\linewidth]{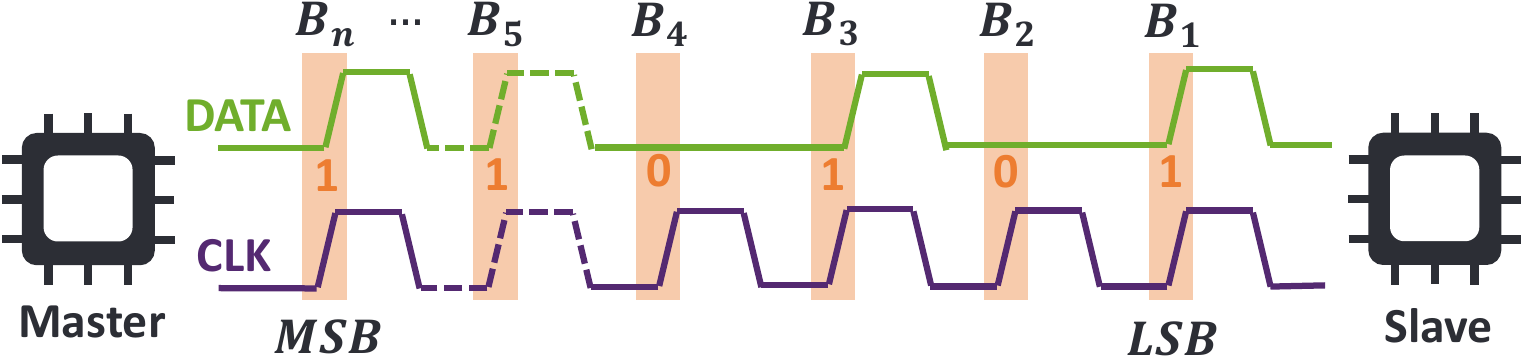}
    \caption{Timing Diagram for Interaction.}
    \label{fig:interaction}
\end{figure}

\textbf{Synchronous Communication Interface.}
Unlike MCUs, standard FPGAs lack native peripheral modules, such as I2C and UART. We implement a bidirectional four-wire interface using precise timing control. As shown in Fig.~\ref{fig:interaction}, the master provides a CLK signal while the slave samples data on rising edges. For FPGA configuration, the MCU acts as master; for downlink data, the FPGA initiates communication with MCU decoding via I/O edge detection.

\textbf{Interaction Frame Structure.}
To further enhance communication reliability, we have customized the interaction frame format, which consists of four fields: a 24-bit preamble (\textit{0xE256E2}) with strong autocorrelation properties for synchronization, an 8-bit ID field for device identification, an 8-bit Type field indicating the data category (e.g., sampling rate, sample count, or symbol duration), and a max 128-bit data field carrying the actual payload. This structured design ensures robust and efficient communication between the MCU and FPGA.

\subsubsection{Parameterized Control Plane}\leavevmode

Building upon this interaction channel, \sysname introduces a control plane that abstracts complex hardware-level peripheral configurations into a unified semantic space. At the heart of this plane is a \textit{Parameter Interpreter} that utilizes macro definitions to translate high-level user inputs into specific hardware commands, enabling the preprocessing of configurable parameters and operational states. Additionally, we develop translation functions to interpret these parameters and transmit them to the hardware interfaces. The user can ultimately control all \sysname behaviors directly through the host computer connected to the MCU.

\begin{figure}[t]
    \centering
    \includegraphics[width=0.98\linewidth]{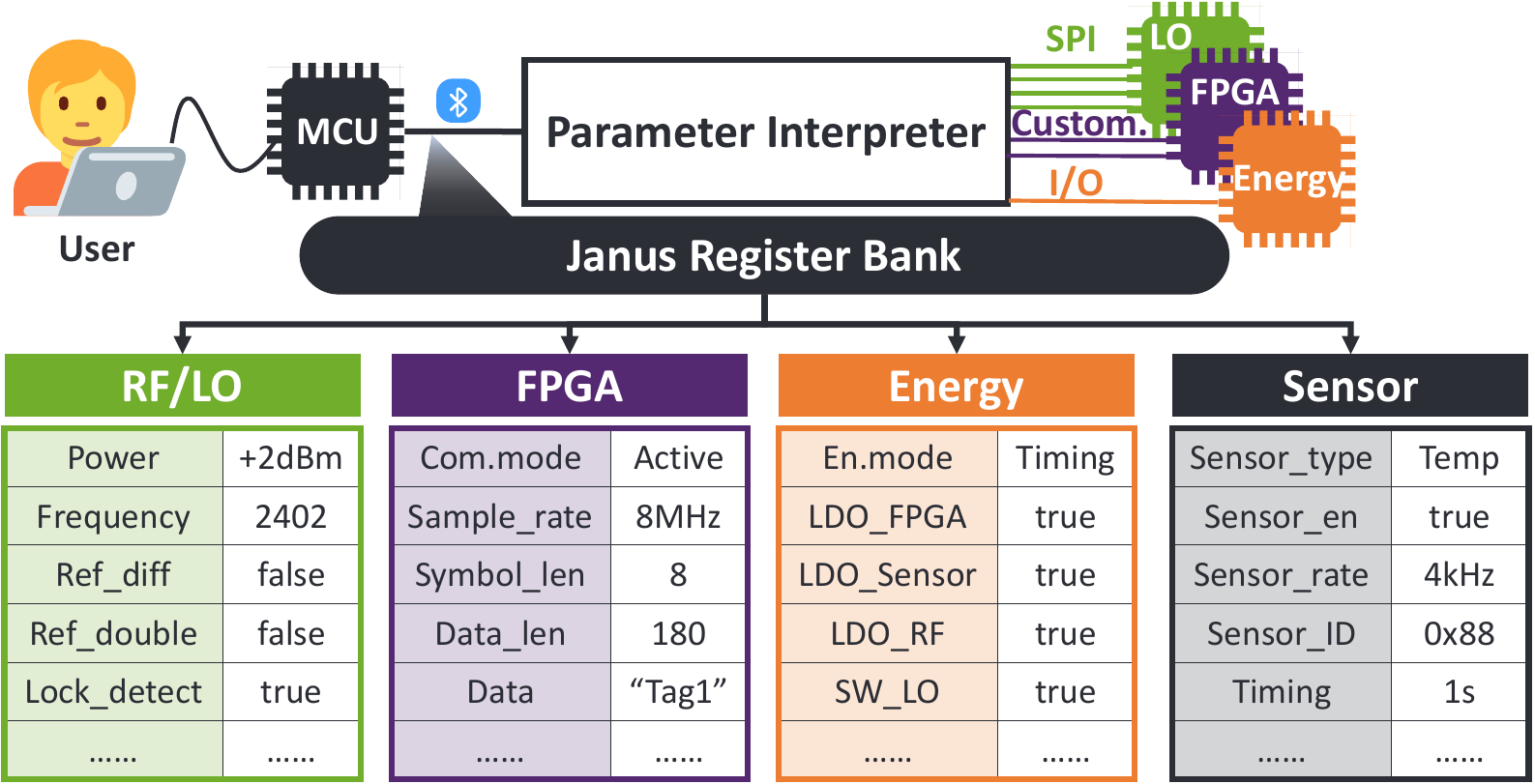}
    \caption{Parameter Configuration Process. Configuration parameters within the MCU are updated via the host UART interface and subsequently translated into hardware peripheral commands by a parameter interpreter module.}
    \label{fig:control_plane}
\end{figure}

Fig.~\ref{fig:control_plane} demonstrates a configuration workflow for active BLE beacon transmission. Users modify the \textit{\sysname Register Bank} to define key parameters, such as setting the transmission power to $+2dBm$ on Channel 37 ($2.402$ GHz) with a ``Tag1'' payload identifier. Additionally, the configuration enables the temperature sensor for periodic data transmission.

\subsection{Power management Plane}\label{Sec:3.3}

To transcend the energy limitations inherent in standalone A-IoT operation, \sysname incorporates a dedicated energy management plane designed to bridge the gap between volatile ambient scavenging and stable system execution. 

\begin{figure}[t]
    \centering
    \includegraphics[width=0.98\linewidth]{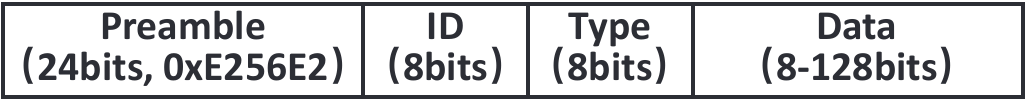}
    \caption{Frame Structure for MCU-FPGA Interaction.}
    \label{fig:frame}
\end{figure} 

\subsubsection{Energy Harvesting}\leavevmode

A-IoT devices operate in diverse environments (indoor, outdoor, and high-temperature scenarios), requiring multi-source energy harvesting capabilities. For indoor applications (e.g., automated warehouses), RF energy harvesting becomes essential due to limited solar/thermal availability, while outdoor scenarios (e.g., smart agriculture) can leverage solar energy. Thus, \sysname incorporates extensible energy harvesting interfaces to accommodate various sources.

As shown in Fig.~\ref{fig:overview}, the harvesting subsystem comprises a multi-source energy harvesting module and a power management IC BQ25570\cite{BQ25570}. This design achieves $93\%$ power conversion efficiency with $110mA$ peak output current (sufficient for PLL operation) while maintaining merely $488nA$ quiescent current. Our prototype (~\S~\ref{Sec:5.1}~) demonstrates successful RF and solar energy harvesting. An optional battery interface further extends deployment flexibility.

%
\subsubsection{Power Domain Management}\leavevmode

\sysname is designed for energy-constrained A-IoT devices, requiring maximized energy efficiency. We employ multiple low-quiescent-current voltage regulators (TPS782~\cite{TPS782} and TLV702~\cite{TLV702}) and power-distribution switches (TPS2041B~\cite{TPS2041B}) to independently control each power domain. The MCU, serving as \sysname’s controller, operates directly once sufficient energy is available. We configure an \textit{Energy Register Bank} on the MCU to manage the operational states of other power domains.

%
\subsubsection{Operation Modes}\leavevmode

\sysname leverages energy harvesting and fine-grained power management to support two distinct modes:

\textbf{Duty Cycle Mode.} Tailored for intermittent tasks, this mode ensures ultra-low power consumption via periodic sleep-wake cycles. The MCU enters the LPM3.5 deep-sleep state, retaining only RTC functionality, and is reactivated via RTC-triggered interrupts.

\textbf{Exhaustion Mode.} Designed for immediate response, this mode activates the system once harvested energy exceeds a preset threshold. \sysname executes communication tasks until energy depletion, after which it automatically reverts to the energy harvesting state.

\section{Putting Everything Together}
Overall, the SlimSDR prototype system follows a three-layer architecture as illustrated in the figure. Each layer consists of a $5\times5cm$ PCB: the bottom layer is the ``Active PLL Board'', the middle layer is the ``Core Board'', and the top layer is the ``Harvesting Board''. These three PCBs are mechanically stacked and secured using brass standoffs, with interconnections established via data lines and coaxial cables.

\textbf{Active PLL Board.}
This four-layer PCB centers around the fractional-N PLL ADF4355~\cite{ADF4355}, featuring voltage regulators and a $20MHz$ crystal oscillator for reference clock generation. The board integrates power, data, and control interfaces through a 7-pin connector, while transmitting the generated carrier signal to the Core Board via SMA port and coaxial cable.

\textbf{Core Board.}
Serving as the most complex and functional PCB in SlimSDR, this 4-layer board integrates a Gowin GW1N~\cite{gowin}, an MSP430FR5969~\cite{msp430fr5969}, and a variety of interfaces. For reception, it utilizes an LT5534 envelope detector~\cite{LT5534} paired with a TS3021 comparator~\cite{TS3021}, achieving a $-60dBm$ sensitivity comparable to SoTA OOK receivers. It also includes multiple voltage regulators and power-distribution switches for efficient power management. The Tx/Rx RF front-end is implemented on this board.

\textbf{Harvesting Board.}
This two-layer PCB specializes in energy harvesting, employing a single-stage rectifier with SMS7630 Schottky diodes~\cite{SMS7630}, capacitors, and impedance-matching circuitry. The BQ25570-based~\cite{BQ25570} energy management module stores harvested energy in a $90mF$ AVX BestCap supercapacitor~\cite{AVXbestcap}, chosen for its low leakage characteristics, while providing interfaces for supplementary energy sources.

\section{Evaluation}\label{Sec:5}
\subsection{Microbenchmarks and Specifications}\label{Sec:5.1}
\subsubsection{Power Consumption}\leavevmode

\textbf{IC Power Consumption.}
We first evaluate the IC power consumption of the \sysname. The ASIC is implemented using the SMIC $40nm$ $1.1V$ ULP process, with analog and digital components synthesized via Cadence Virtuoso~\cite{Virtuoso} and Synopsys DC~\cite{DCUltra}, respectively. In the RF front-end, the fully passive envelope detector employs a pseudo-balun structure. The comparator ($0.41\mu W$ at $2MHz$), RF switch ($5.3\mu W$), and $40MHz$ ring oscillator-based PLL ($26.2\mu W$) constitute the core power consumers. The SD-baseband, capable of supporting the PHY layer of BLE, 802.11AMP, and 3GPP A-IoT, consumes $29.87\mu W$, $56.77\mu W$, and $29.06\mu W$, respectively. The control plane for parameter configuration remains inactive during operation and consumes only $0.21\mu W$. Furthermore, as referenced in~\cite{jo2023135fs}, the PLL consumes $17.1mW$ during active communication. Therefore, \sysname exhibits minimum power overheads of $61.18\mu W$ in passive mode and $17.2mW$ in active mode.

\begin{figure*}
\centering
    \begin{minipage}[t]{0.37\linewidth}
        \setlength{\belowcaptionskip}{-8pt}
        \centerline{\includegraphics[width=1\linewidth]{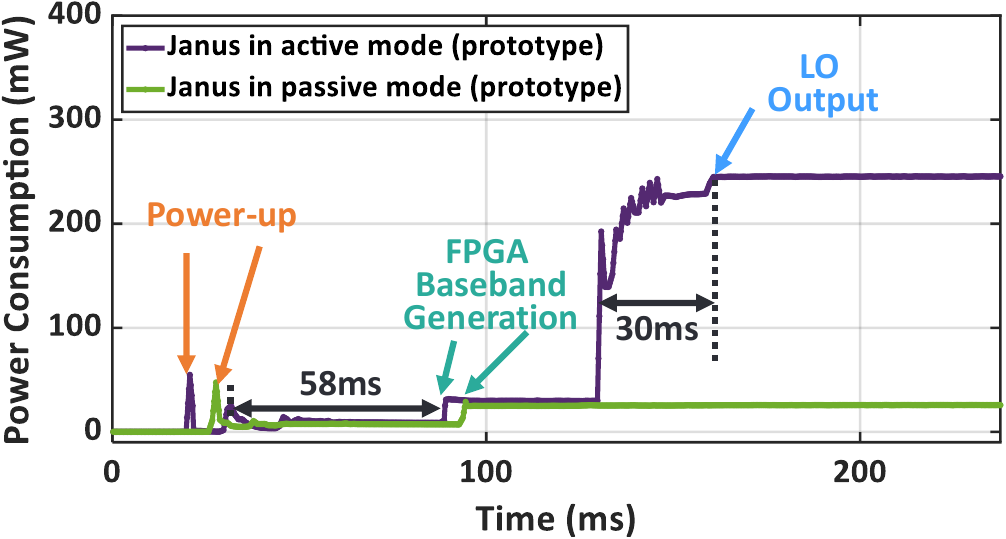}}
        \caption{Power consumption timing diagram of \sysname prototype power-up to operating in different modes.}
        \label{fig:power_consumption}
    \end{minipage} 
    \hspace{2pt}
    \begin{minipage}[t]{0.6\linewidth}
        \setlength{\belowcaptionskip}{-8pt}
        \centerline{\includegraphics[width=1\linewidth]{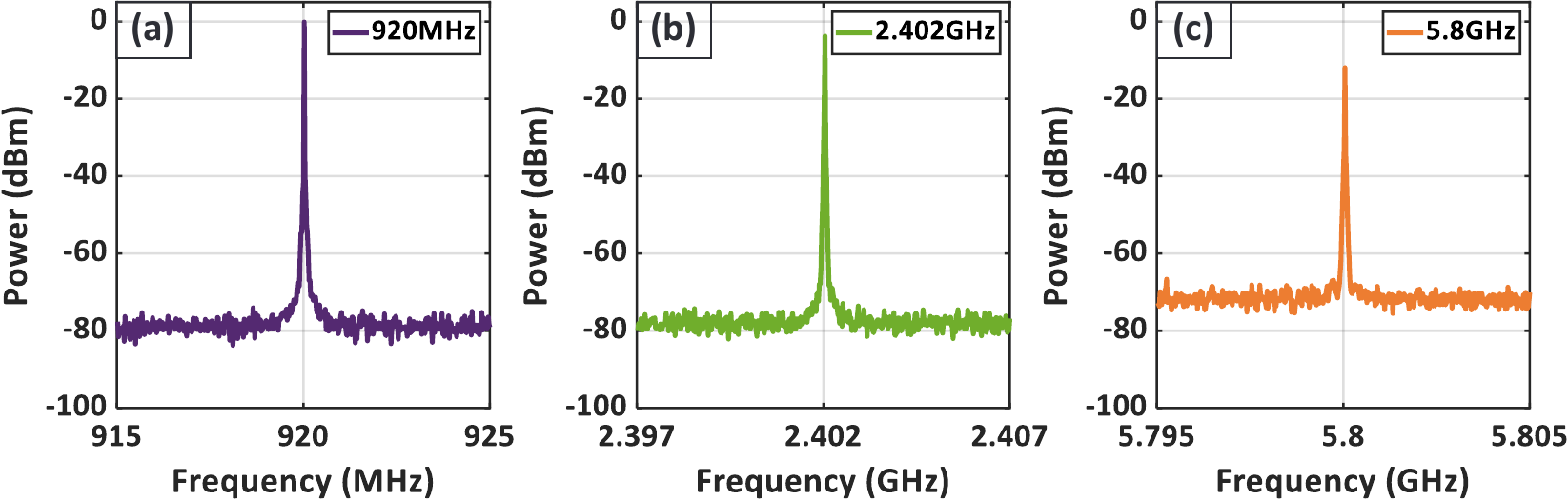}}
        \caption{The spectrograms of the local oscillator output are configured at different frequencies. The signals are precisely centered at the target frequencies without harmonic distortion or spectral spurs.}
        \label{fig:single_tone}
    \end{minipage}
\end{figure*}

\textbf{Prototype Power Consumption.}
The \sysname prototype integrates a low-power FPGA, MCU, fractional-N PLL, voltage regulators, and discrete passive components. Using an Otii Arc energy analyzer~\cite{Otii}, we measure its power consumption in both active and passive modes, as shown in Fig.~\ref{fig:power_consumption}. In passive mode, the MCU initiates first, enabling the FPGA's power domain before FPGA activation and subsequent MCU-FPGA configuration. During active mode, the MCU activates and configures the PLL after FPGA initialization. Measurements show that the power consumption is $24.8mW$ in passive and $249mW$ in active. 
\begin{figure*}[t]
    \centering
    \includegraphics[width=0.8\linewidth]{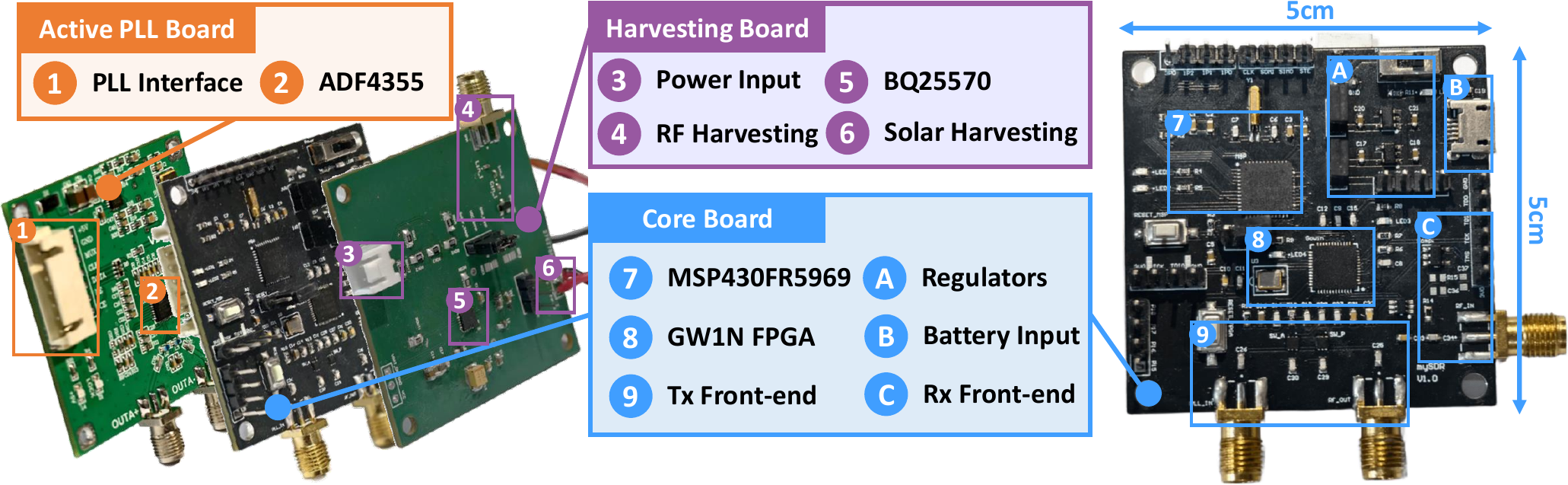}
    \caption{Hardware Prototype of \sysname.}
    \label{fig:hardware}
\end{figure*}

\textbf{Prototype vs ASIC realization.}
While the prototype facilitates functional verification, the ultimate objective is to realize \sysname as a fully integrated SoC in the future. As detailed in Table.~\ref{table:comparsion_power}, the ASIC implementation reduces power consumption by two to three orders of magnitude compared to the prototype. This disparity primarily stems from the overhead of discrete active components (e.g., MCU, FPGA, PLL) and board-level inefficiencies, such as parasitic effects, interface redundancy, static power accumulation and conversion losses. Furthermore, the RTL-based control logic and SPI interfaces enable a seamless migration to ASIC design, thereby preserving the parametric configurability of \sysname.

\textbf{Sleep Mode.}
In sleep mode, with the MCU operating in LPM3.5, the system consumes $18\mu W$. This represents a $1.7\times$ reduction compared to the $30\mu W$ consumption of TinySDR~\cite{tinysdr}.

\subsubsection{Switching Delays}\leavevmode
We evaluate the latency of \sysname when transitioning from sleep mode to normal operational mode. The measured switching delay from sleep mode to active radio operation is $88.8ms$, which includes $0.3ms$ for the MCU to exit LPM3.5 mode, $0.5ms$ for regulator enablement, $58ms$ for FPGA reinitialization and configuration, and $30ms$ for PLL stabilization. In contrast, the switching latency from sleep mode to backscatter mode operation is $58.8ms$, as it excludes the PLL power-up process. To ensure reliable operation, the \sysname prototype incorporates a $100ms$ timer on the MCU, allowing sufficient time for FPGA reinitialization and configuration. The corresponding timing diagram is illustrated in Fig.~\ref{fig:power_consumption}.

\subsubsection{Local Oscillator Performance}\leavevmode
\sysname is designed to realize active transmission across a wide frequency band. We evaluate the performance of the local oscillator at different output frequencies. By configuring the frequency synthesizer via the SPI interface from MCU, we set the output to 0dBm at three distinct frequencies: $920MHz$, $2.402GHz$, and $5.8GHz$. The output is connected to an N9322C Spectrum Analyzer via a coaxial cable, with a resolution bandwidth of $1kHz$ and a sweep bandwidth of $10MHz$. As shown in Fig.~\ref{fig:single_tone}, the results demonstrate that the carrier signals are precisely centered at the target frequencies without harmonic distortion or spectral spurs. However, as the output frequency increased, higher attenuation is observed, leading to a reduction in output power below the ideal configured value.

%
\subsubsection{Energy Harvesting}\leavevmode
We assess the operational feasibility of \sysname using harvested energy. Following the BQ25570 datasheet specifications with $10k\Omega$ input impedance, we employ the same matching load to measure input power from diverse energy sources.

\begin{figure*}
\centering
    \begin{minipage}[t]{0.48\linewidth}
        \setlength{\belowcaptionskip}{-8pt}
        \centerline{\includegraphics[width=1\linewidth]{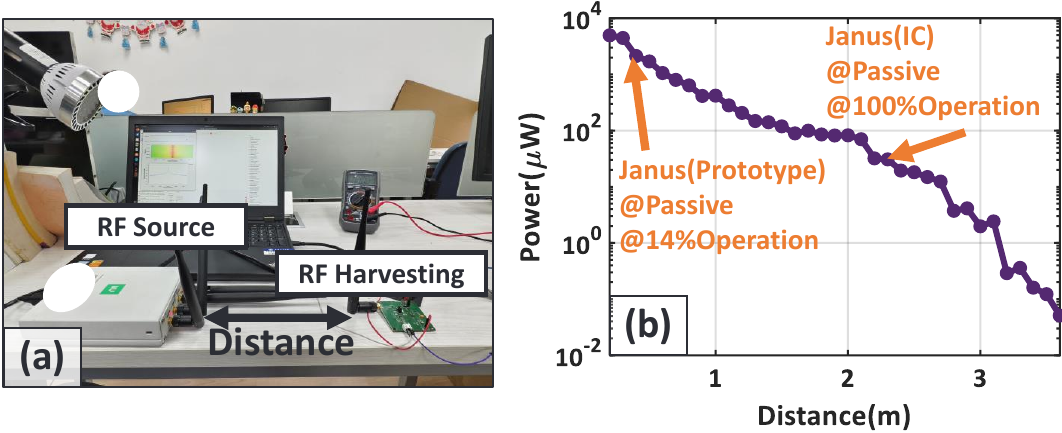}}
        \caption{Experiments for RF Energy Harvesting. (a) Experimental setup. (b) Curve of harvesting energy as a function of distance.}
        \label{fig:exp_harvesting_rf}
    \end{minipage} 
    \hspace{0pt}
    \begin{minipage}[t]{0.48\linewidth}
        \setlength{\belowcaptionskip}{-8pt}
        \centerline{\includegraphics[width=1\linewidth]{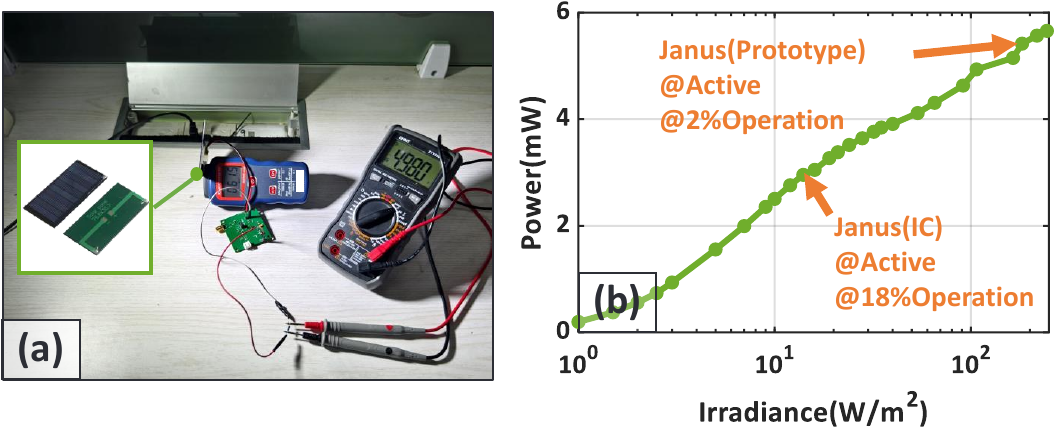}}
        \caption{Experiments for Solar/Light Energy Harvesting. (a) Experimental setup. (b) Curve of harvesting energy as a function of irradiance.}
        \label{fig:exp_harvesting_solar}
    \end{minipage}
\end{figure*}

\textbf{RF Energy Harvesting.}
We employ a USRP B210 equipped with an MWPA0050G04 RF amplifier ($36dBm$ EIRP with a $3dBi$ antenna at $915MHz$). As shown in Figure~\ref{fig:exp_harvesting_rf}, the single-stage rectifier of the \textit{Tiga} tag prototype can harvest sufficient power to support passive operation (14\% duty cycle) at a 0.3-meter distance, while the ASIC-based \sysname can operate in real time at a distance of 1.2 meters.
\begin{table}[t]
    \centering
    \caption{\centering Power consumption comparison \\ between SoTAs.}
    \includegraphics[width=0.98\linewidth]{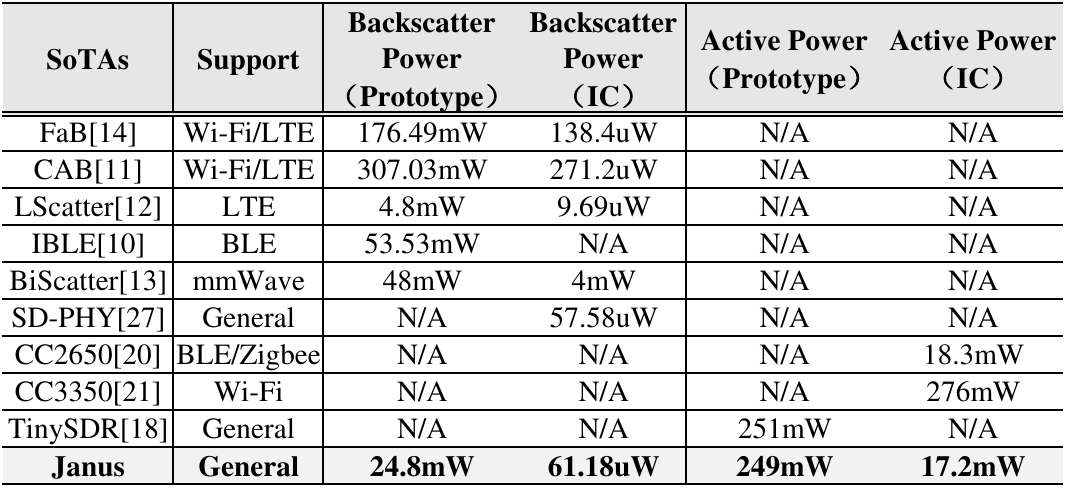}
    \label{table:comparsion_power}
\end{table}

\textbf{Solar/Light Energy Harvesting.}
To evaluate solar energy availability for \sysname, we set up an indoor experiment using a variable-power LED as the light source. A $53mm\times30mm$ mini solar panel~\cite{Solarpanel} is employed for energy harvesting, while a solar power meter SM206E-Solar~\cite{SM206E} monitors the irradiance level. As shown in Fig.~\ref{fig:exp_harvesting_solar}, when irradiance reaches $164W/m^2$, \sysname can operate in active mode with a $2\%$ duty cycle. When irradiance is reduced to $16W/m^2$, the ASIC-based \sysname can operate in active mode with an $18\%$ duty cycle. Since sunlight is attenuated by the atmosphere, the maximum normal surface irradiance at sea level on a clear day is approximately $1000W/m^2$~\cite{wikisolar}. Therefore, outdoor deployment of \sysname can readily support its operation.

\subsection{\sysname in Action}\label{Sec:5.2}
In this section, we deploy a prototype \sysname device for case studies involving representative standardization organizations: IEEE 802.11AMP, 3GPP A-IoT and Bluetooth SIG. 
Fundamentally, the radio hardware functions as an arbitrary waveform RF transceiver. We focus on the physical layer, specifically on modulation for transmission and demodulation for reception; the implementation of higher protocol layers is reserved for our future work.
In the communication performance evaluation, we employ a LiPo battery as the stable energy source for \sysname.

\subsubsection{802.11AMP with \sysname}\leavevmode

\textbf{802.11AMP Primer.} The IEEE 802.11AMP standard~\cite{80211AMP} is designed to facilitate ultra-low-power, A-IoT operations within WLAN ecosystems. This architecture defines a hybrid framework for both backscatter and active devices operating across the $2.4GHz$ and Sub-$1GHz$ frequency bands. In the downlink, the standard synthesizes OOK using OFDM symbols to support data rates of $250Kbps$ and $1Mbps$. Conversely, the uplink employs direct OOK modulation with Manchester encoding. While both backscatter and active STAs support uplink rates of $250Kbps$ and $1Mbps$, a higher data rate of $4Mbps$ is reserved exclusively for active STAs.

\begin{figure*}
\centering
    \begin{minipage}[t]{0.44\linewidth}
        \setlength{\belowcaptionskip}{-8pt}
        \centerline{\includegraphics[width=1\linewidth]{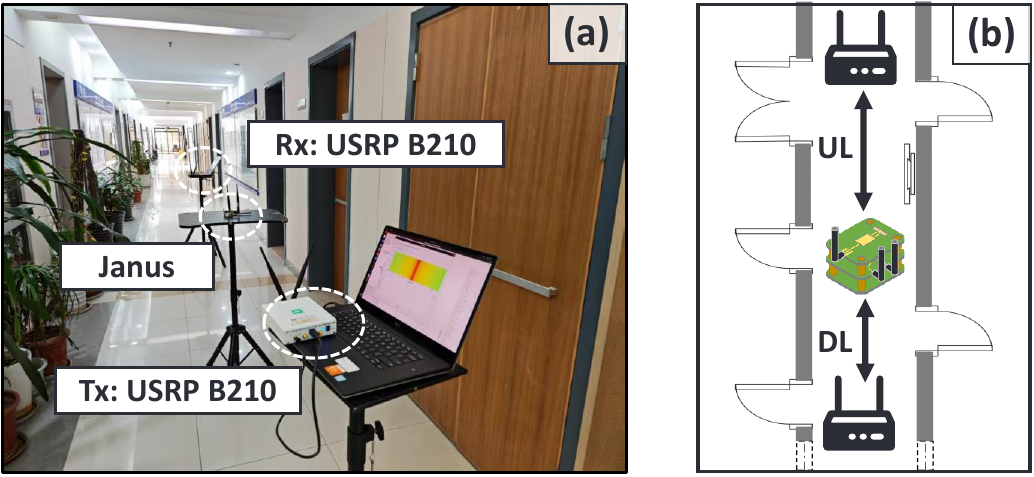}}
        \caption{\sysname realizes the PHY layer of 802.11AMP and 3GPP A-IoT. (a) The experiments are conducted in the corridor. (b) Floor plan.}
        \label{fig:exp_80211AMP_setup}
    \end{minipage} 
    \hspace{2pt}
    \begin{minipage}[t]{0.54\linewidth}
        \setlength{\belowcaptionskip}{-8pt}
        \centerline{\includegraphics[width=1\linewidth]{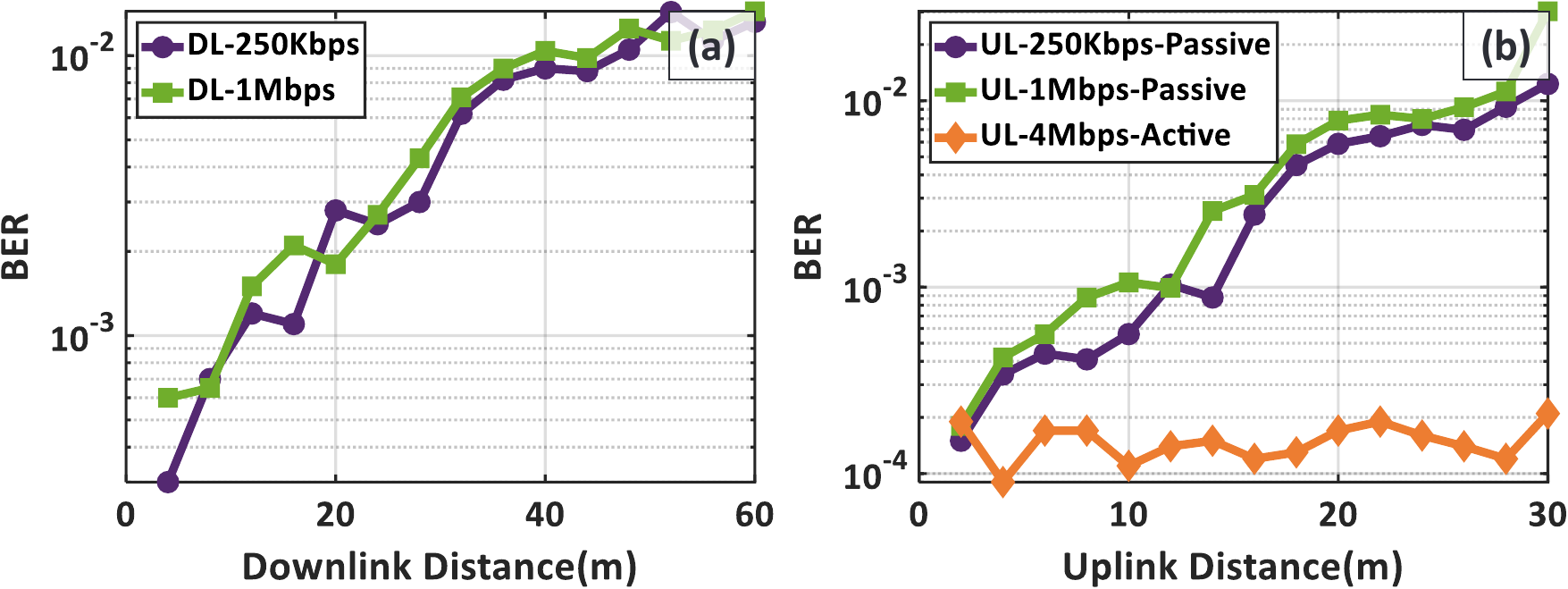}}
        \caption{Performance comparison of the \sysname in uplink and downlink under 802.11AMP. (a) BER at different data rates in the downlink. (b) BER at different data rates in the uplink.}
        \label{fig:exp_80211AMP_result}
    \end{minipage}
\end{figure*}

\textbf{Results.} To evaluate the performance of \sysname within this framework, we configure the device for OOK modulation and deploy it in a corridor environment, as illustrated in Fig.~\ref{fig:exp_80211AMP_setup}(a) and (b). Two USRP B210s are employed as the transmitter and receiver, respectively. The transmitter operated at a center frequency of $2.412GHz$, emitting OOK data signals alongside a single tone at $15dBm$. In the passive backscatter mode, the \sysname is positioned at a standoff distance of $0.5m$ from the transmitter. As shown in Fig.~\ref{fig:exp_80211AMP_result}, experimental results demonstrate that the \sysname OOK receiver maintains BER below $1\%$ at downlink data rates of both $250Kbps$ and $1Mbps$ over communication distances up to $44m$. In the uplink, passive backscatter communication achieves a BER of less than $1\%$ within a range of $28m$. In active mode, benefiting from the local oscillator, the BER remains consistently around 0.01\%.

\begin{figure*}
\centering
    \begin{minipage}[t]{0.49\linewidth}
        \setlength{\belowcaptionskip}{-8pt}
        \centerline{\includegraphics[width=1\linewidth]{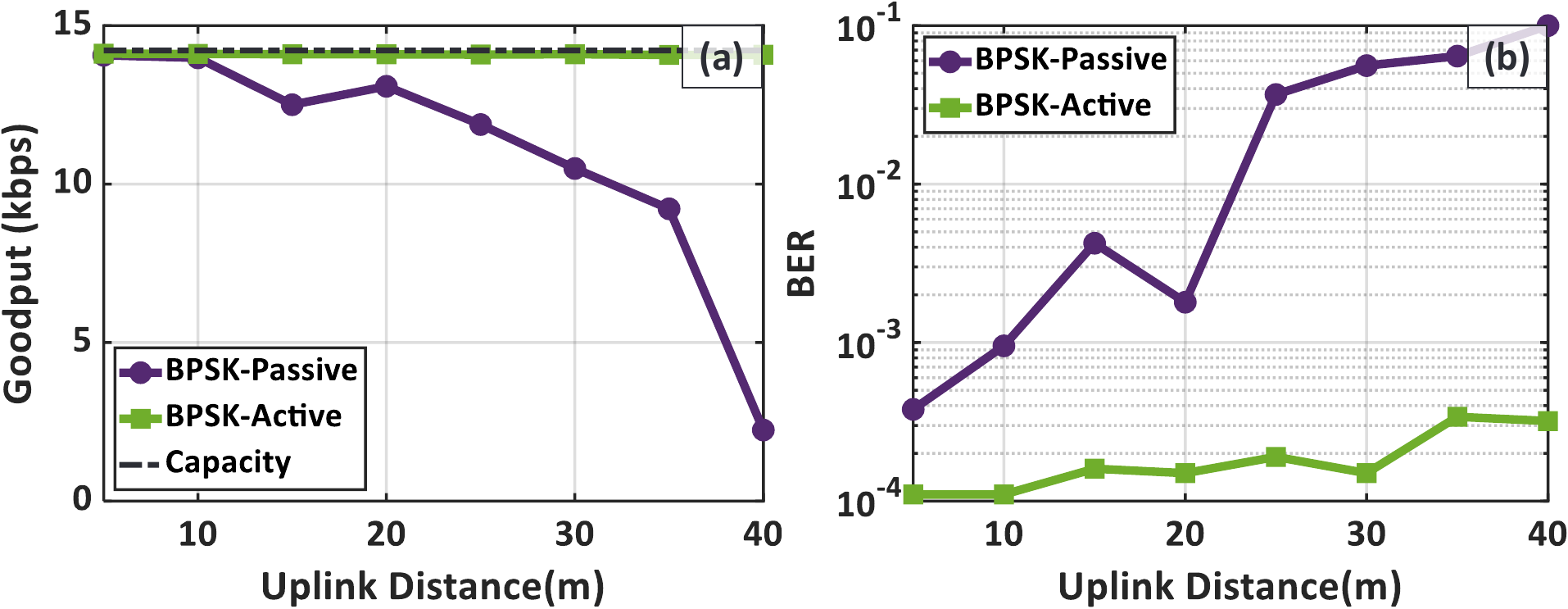}}
        \caption{Performance comparison of the \sysname configured with BPSK modulation under the 3GPP A-IoT protocol. (a) Goodput. (b) BER.}
        \label{fig:exp_3gpp_BPSK_result}
    \end{minipage} 
    \hspace{2pt}
    \begin{minipage}[t]{0.49\linewidth}
        \setlength{\belowcaptionskip}{-8pt}
        \centerline{\includegraphics[width=1\linewidth]{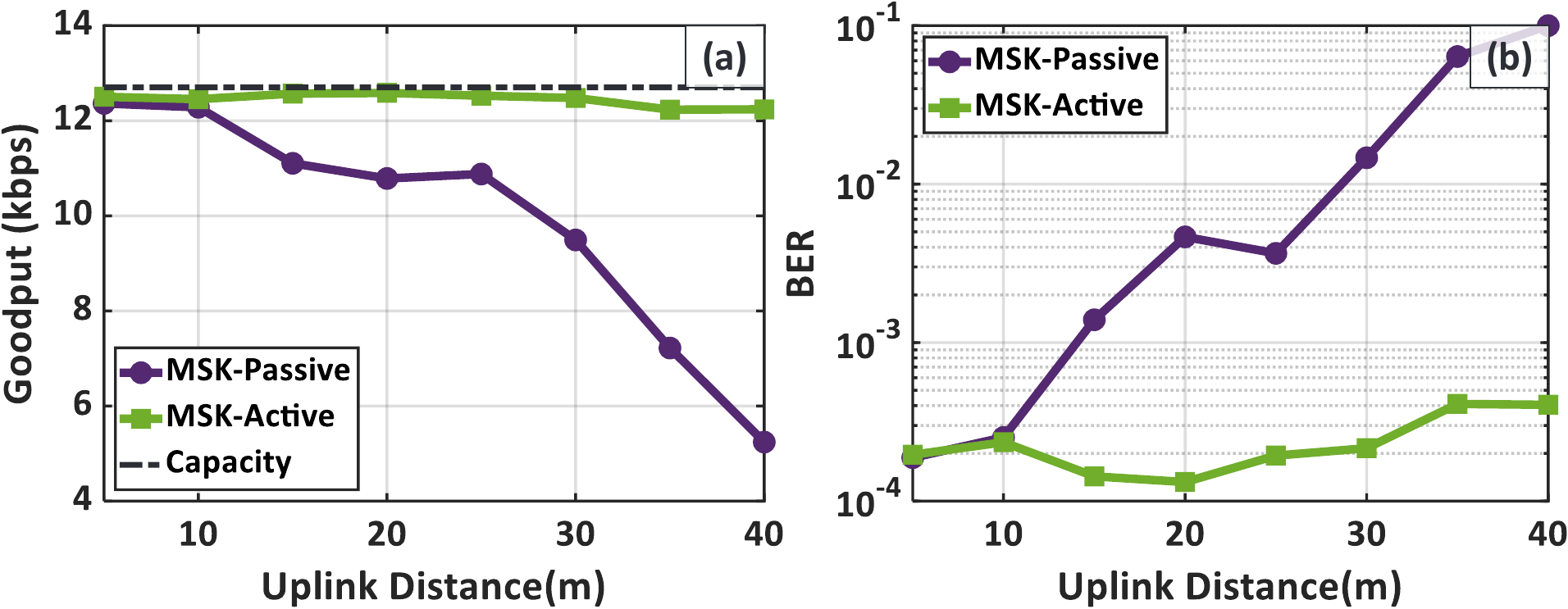}}
        \caption{Performance comparison of the \sysname configured with MSK modulation under the 3GPP A-IoT protocol. (a) Goodput. (b) BER.}
        \label{fig:exp_3gpp_FSK_result}
    \end{minipage}
\end{figure*}
\begin{figure*}
\centering
    \begin{minipage}[t]{0.26\linewidth}
        \setlength{\belowcaptionskip}{-8pt}
        \centerline{\includegraphics[width=1\linewidth]{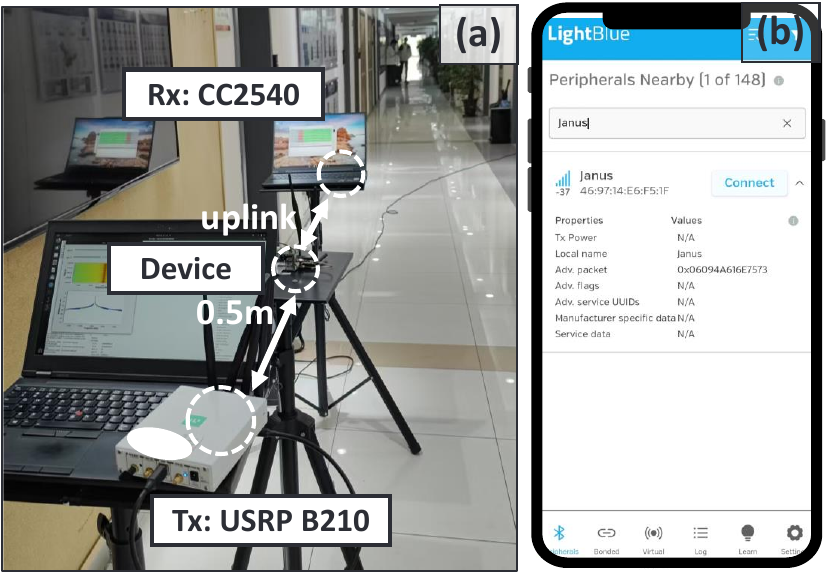}}
        \caption{Comparative experiments. The experiments are conducted in the corridor.}
        \label{fig:exp_ble_setup}
    \end{minipage} 
    \hspace{0pt}
    \begin{minipage}[t]{0.31\linewidth}
        \setlength{\belowcaptionskip}{-8pt}
        \centerline{\includegraphics[width=1\linewidth]{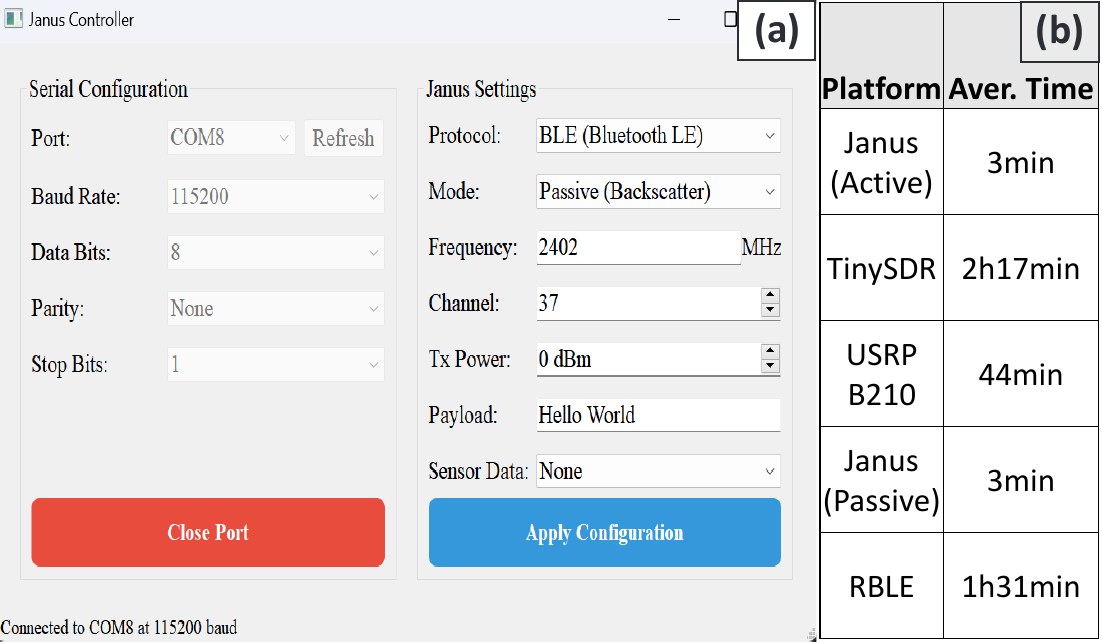}}
        \caption{Configuration Method. (a) Host interface of the \sysname. (b) Configuration efficiency.}
        \label{fig:host_computer}
    \end{minipage}
    \hspace{0pt}
    \begin{minipage}[t]{0.4\linewidth}
        \setlength{\belowcaptionskip}{-8pt}
        \centerline{\includegraphics[width=1\linewidth]{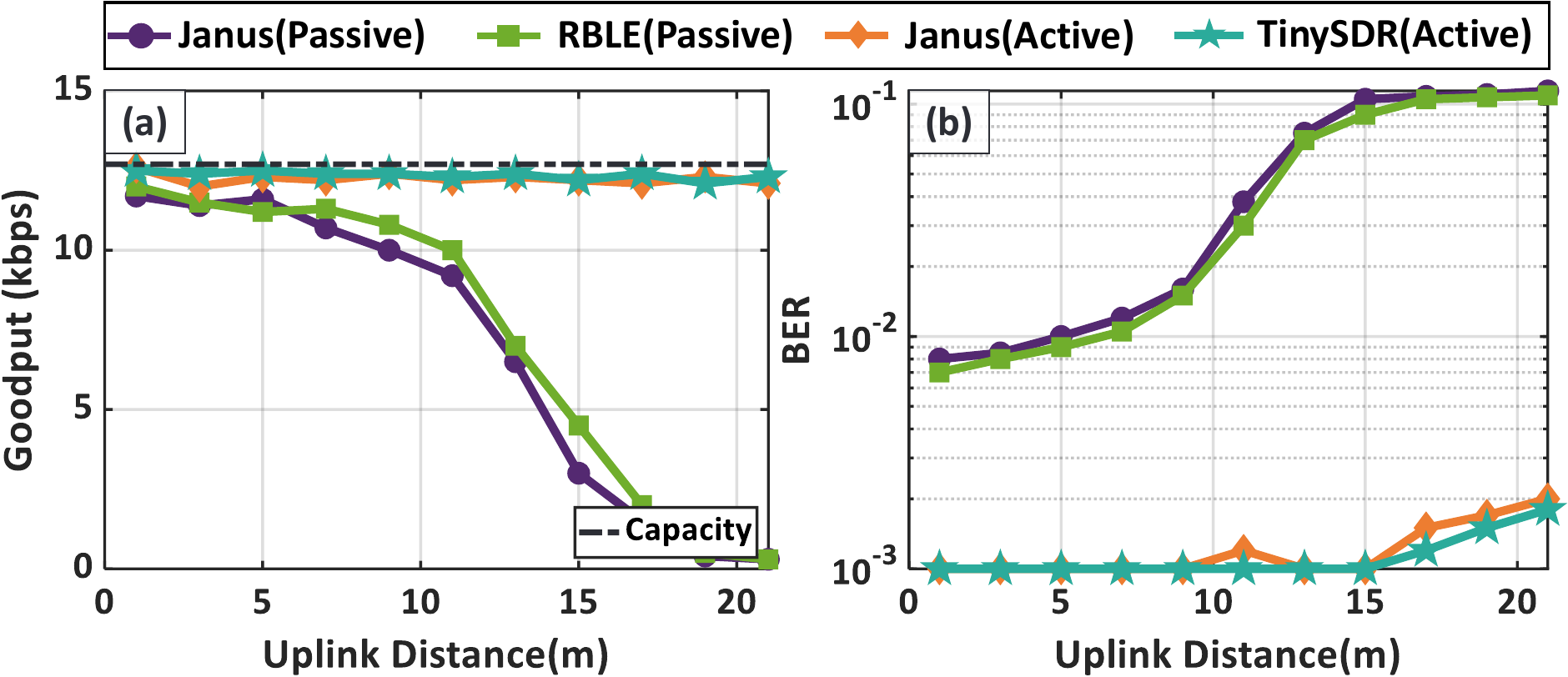}}
        \caption{Comparison of the performance of \sysname and dedicated devices. (a) BLE goodput. (c) BER.}
        \label{fig:exp_ble_result}
    \end{minipage}
\end{figure*}

\subsubsection{3GPP A-IoT with \sysname}\leavevmode

\textbf{3GPP A-IoT Primer.} The 3GPP A-IoT specification~\cite{3gpp_tr38769} establishes a harmonized air interface for ultra-low-complexity, battery-free devices to ensure coexistence with 5G NR. For the downlink, the standard utilizes OFDM-synthesized OOK; conversely, the uplink prioritizes energy efficiency by supporting backscatter or low-power active transmissions using modulation schemes such as OOK, BPSK, and BFSK.

\textbf{Results.} We conduct experiments in a corridor environment with the transmitter operating in 3GPP Band 3. Given the modulation alignment with 802.11AMP, we evaluate the \sysname to operate exclusively in BPSK and BFSK modes in uplink. As defined in the 3GPP A-IoT technical specifications, BFSK is implemented using Minimum Shift Keying (MSK)—a continuous-phase modulation with an index of 0.5. Accordingly, we configure the \sysname such that the frequency separation between symbols `0' and `1' corresponds to half the data rate. The experimental results across varying distances are depicted in Fig.~\ref{fig:exp_3gpp_BPSK_result} and Fig.~\ref{fig:exp_3gpp_FSK_result}. In active mode, both BPSK and MSK demonstrate robust transmission, maintaining a BER below 0.1\% and yielding goodput approaching theoretical capacity. In backscatter mode, however, performance naturally degraded with distance due to the severe attenuation inherent in the round-trip backscatter channel.

\subsubsection{Comparision with dedicated devices.}\leavevmode

\textbf{Setup.} We benchmark \sysname against RBLE for backscatter communication~\cite{RBLE}, and against TinySDR~\cite{tinysdr} and USRP B210~\cite{USRPB210} for active transmission. As depicted in Fig.~\ref{fig:exp_ble_setup} (a), the experiments are conducted in a corridor environment where each device broadcasts valid BLE advertising packets on Channel 37. A TI CC2540 receiver~\cite{CC2540} captured the traffic using a Smart Packet Sniffer~\cite{packetsniffer}. For the backscatter configuration, the distance between the excitation source and the tag is fixed at $0.5m$.
To evaluate configuration efficiency, we recruit ten Ph.D. students with relevant domain knowledge and engineering expertise. Each participant is tasked with configuring every platform independently, and the average configuration time for each platform is recorded. This study involved no ethical concerns.

\textbf{Results.} We first evaluate configuration efficiency. Unlike RBLE and TinySDR, which require time-consuming Verilog synthesis, or the USRP B210, which relies on Python-UHD processing, \sysname supports direct parameter updates via a host UART interface. As shown in Fig.~\ref{fig:host_computer}, this software-defined architecture allows \sysname to complete configuration in approximately 3 minutes, significantly outperforming dedicated devices. 
This parameterized approach achieves a $14.6\times$ to $45.6\times$ speedup in configuration efficiency compared to traditional Python-UHD and Verilog-based methods.
Fig.~\ref{fig:exp_ble_result} confirms that \sysname achieves communication reliability comparable to that of the dedicated hardware platforms.

\subsubsection{Ambient IoT Application with \sysname}\leavevmode

\textbf{Setup.} We evaluate the performance of \sysname in an A-IoT scenario under two distinct conditions: an outdoor environment with an irradiance of $524 W/m^2$ for the BLE active mode, and an indoor setting using a $915 MHz$ RF source ($33 dBm$ EIRP) at a distance of $0.1 m$ for the passive mode.

\begin{figure*}
\centering
    \begin{minipage}[t]{0.39\linewidth}
        \setlength{\belowcaptionskip}{-8pt}
        \centerline{\includegraphics[width=1\linewidth]{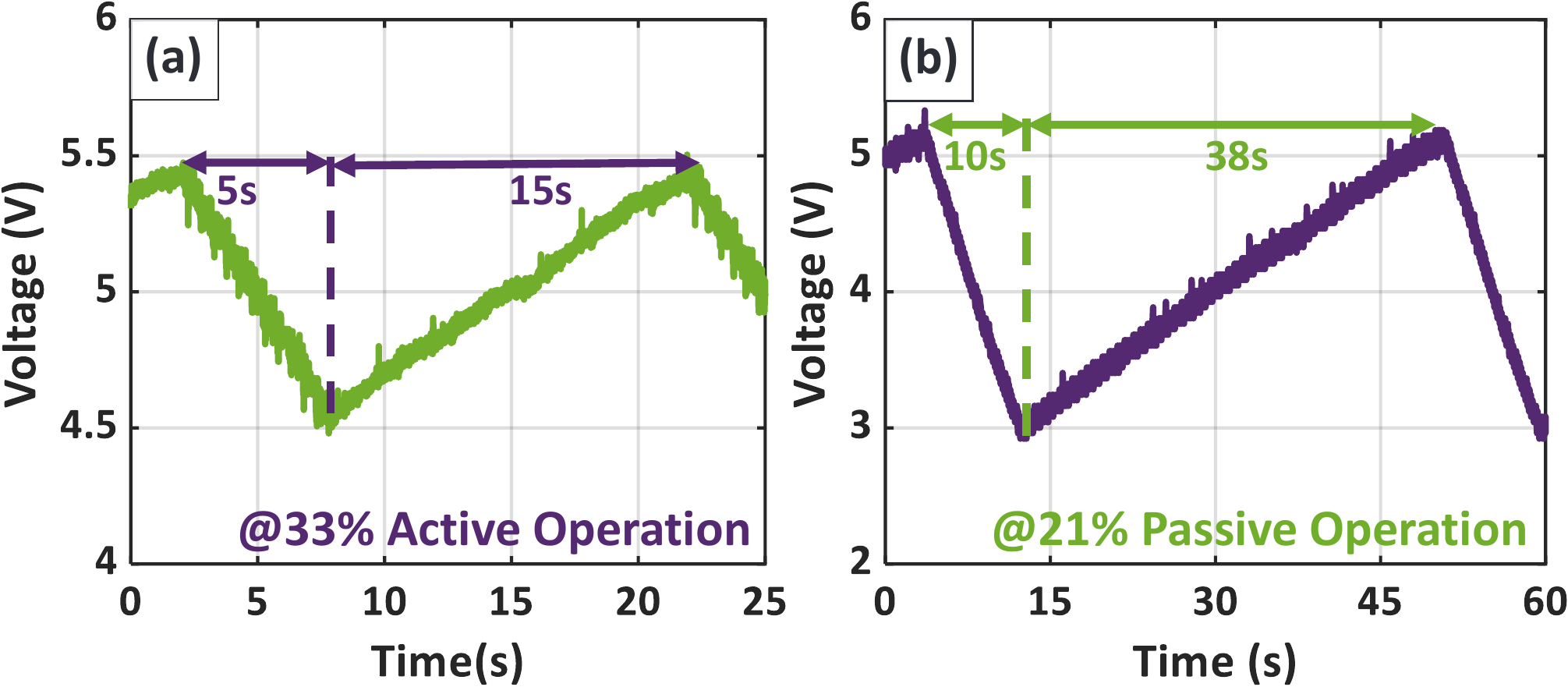}}
        \caption{Voltage profiles of the storage capacitor in \sysname during energy harvesting. (a) Solar harvesting. (b) RF harvesting.}
        \label{fig:exp_harvesting}
    \end{minipage} 
    \hspace{2pt}
    \begin{minipage}[t]{0.59\linewidth}
        \setlength{\belowcaptionskip}{-8pt}
        \centerline{\includegraphics[width=1\linewidth]{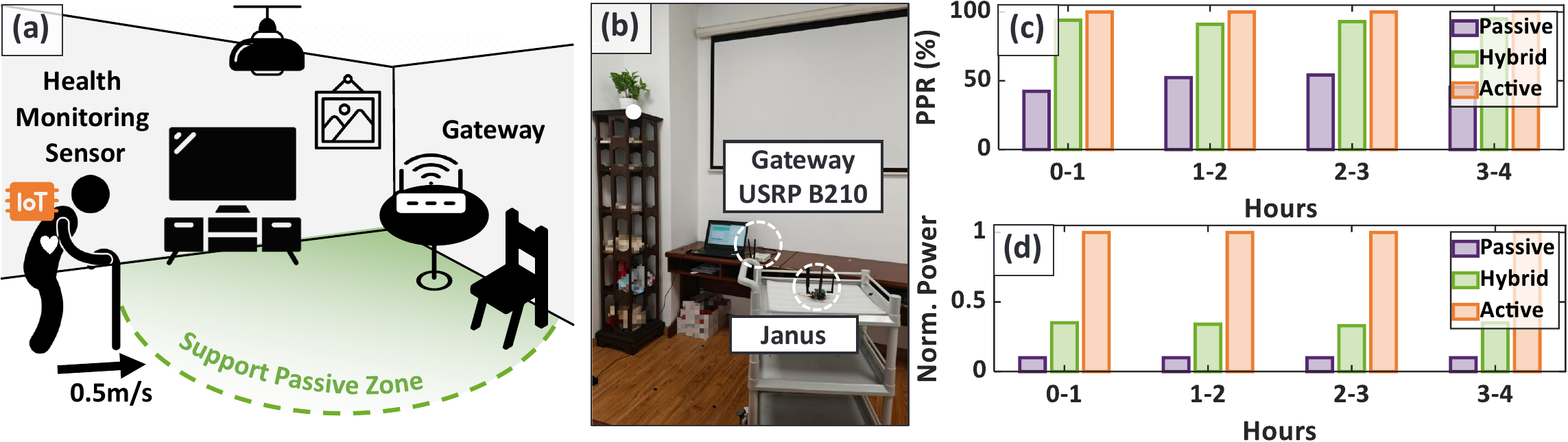}}
        \caption{A proof-of-concept application. (a) A-IoT devices worn on elders support passive communication within the passive zone. (b) Experiment setup. (c) PRR. (d) Normalized power consumption.}
        \label{fig:hybridmode}
    \end{minipage}
\end{figure*}

\textbf{Results.} Figure~\ref{fig:exp_harvesting} illustrates the voltage across the storage capacitor. The results demonstrate that the prototype effectively harvests solar and RF energy to sustain operation, achieving duty cycles of $33\%$ in the active mode and $21\%$ in the passive mode.

\subsubsection{Proof-of-Concept Application of Hybrid Mode}\leavevmode

We demonstrate the \sysname hybrid-mode operation framework through an indoor elderly health monitoring system, a representative A-IoT application~\cite{3GPP3} illustrated in Fig.~\ref{fig:hybridmode}(a). In this scenario, devices proximal to the gateway harvest sufficient energy for passive operation, whereas distant nodes necessitate active mode to ensure reliability. Intuitively, this spatial variation naturally leads to the formation of a passive communication zone.

\textbf{Setup.}
Experimental validation employs a USRP B210 with an output power of $20dBm$ as both the transceiver and excitation source. \sysname transmits uplink BLE beacons on Channel 37 and receives $2.4GHz$ OOK downlink signals. In active mode, the transmission power is set to $6dBm$. Mode switching is driven by uplink signal strength: the gateway triggers a transition to passive mode and commands \sysname via the downlink when the measured RSSI exceeds a threshold of $-35dBm$.

\textbf{Results.}
We conduct a four-hour experiment simulating random indoor movement to evaluate the Packet Reception Ratio (PRR) and normalized power consumption. As shown in Fig.~\ref{fig:hybridmode}(c) and (d), the hybrid mode achieves a PRR comparable to the active mode while consuming only one-third of the power. Although energy consumption is higher than in the purely passive mode, hybrid operation guarantees the link stability essential for critical health monitoring. Furthermore, this platform facilitates the validation of diverse adaptive strategies, including energy- and data-rate-based selection~\cite{kan2023adaptive, ma2022covert, zargari2022energy, kim2020backscatter, hu2016braidio}.

\section{Discussion}
\subsection{Radio Structure.}
Commercial radios~\cite{USRPB210,USRPE310,bladeRF,plutoSDR} typically employ orthogonal I/Q transceivers to support higher-order modulation schemes (e.g., 16QAM, 64QAM) while suppressing image frequency interference. HackRF One~\cite{HackRF} utilizes a superheterodyne architecture with MAX2839 ($2.17-2.74GHz$) and RFFC5072 ($84.375MHz-5.4GHz$) chips to achieve $1MHz-6GHz$ coverage, albeit with increased hardware complexity. 
In contrast, \sysname’s core novelty lies in being the first architecture to natively unify active and passive communication. Disjointed hardware stacking approaches (e.g., Braidio~\cite{hu2016braidio} and Protean~\cite{bakar2022protean} ) suffer reflashing delays and cannot share a unified baseband, as passive mode fundamentally relies on rapid impedance switching while active mode requires a DAC and carrier mixing (Fig.~\ref{fig:rf_frontend}). By exploiting the RF switch’s dual role as an impedance modulator in passive mode and a mixer/DAC in active mode, combined with a custom adapter tailored to our unified RF front-end in baseband, \sysname enables seamless modality transitions on a minimalist hardware substrate.

\subsection{Energy Availability.}
A-IoT devices can harvest energy from RF, solar, thermal, or mechanical sources. RF harvesting suits controlled indoor environments but faces challenges in wireless scenarios due to low power density (typically below $10nW/cm^2$~\cite{kim2014ambient}) and the limitation of forward voltage of diodes in rectifiers (typically $0.34V$~\cite{SMS7630}). Our experiments (\S~\ref{Sec:5}) confirm RF can sustain passive-mode operation at low duty cycles, though active mode requires future ASIC improvements. Solar energy offers higher density (up to $100mW/cm^2$~\cite{green2021solar}) and successfully powers active-passive operation, yet suffers from intermittency due to environmental factors.

\section{Related Work}
\subsection{RF Radio Platform.}
Commercial off-the-shelf radios, such as USRP B210~\cite{USRPB210}, USRP E310~\cite{USRPE310}, BladeRF 2.0~\cite{bladeRF}, and Pluto SDR~\cite{plutoSDR}, employ high-performance transceiver chips and processor SoCs. These solutions provide wide frequency coverage, high bandwidth, high sampling rates, and superior receiver sensitivity. Additionally, open-source platforms like HackRF One~\cite{HackRF}, WARP~\cite{warp}, and $\mu$SDR~\cite{uSDR} adopt similar designs, enabling hardware accessibility for developers. Researchers have also designed or modified existing platforms (e.g., M-Cube~\cite{mcube}, SweepSense~\cite{sweepsense}, Tick~\cite{tick}, OpenMili~\cite{openmili}, SDR-Lite~\cite{jeong2020sdr}, mmFD~\cite{singh2020millimeter}, Whisper~\cite{chakraborty2022whisper}) for specialized radio applications. However, these platforms rely on energy-hungry components primarily targeting base stations or gateways, where energy consumption is not a critical concern.

TinySDR~\cite{tinysdr} represents a pioneering radio platform tailored for IoT endpoints, featuring an architecture optimized for low-power deployment. Nevertheless, the emerging A-IoT paradigm introduces distinct challenges, specifically the necessity for hybrid active-passive communication and multi-source energy harvesting. Furthermore, bridging the gap between hardware-level FPGA synthesis and rapid protocol prototyping remains a key opportunity for enhancing system usability.
SD-PHY~\cite{sdphy} presents a notable case of software-defined backscatter signal modulation in backscatter communication.
Although SD-PHY remains some distance from fully meeting the requirements of next-generation A-IoT devices, it opens a new pathway for \sysname.
\sysname is specifically designed for next-generation A-IoT devices, enabling full operational control through parameterization. 
It rethinks radio platform hardware design by addressing A-IoT requirements, including energy harvesting, hybrid active-passive operation, low power consumption, and user experience.

\subsection{Battery-free Platforms.}
The pursuit of battery-free operation has been a persistent research objective in IoT systems, with primary efforts directed toward minimizing energy consumption in communication, sensing, and actuation while simultaneously enhancing energy harvesting and management efficiency. However, commercial battery-free platforms such as EnOcean EWSSZ~\cite{ewssz}, Wiliot Pixel~\cite{wiliot}, and Everactive Eversensor~\cite{everactive} are constrained by proprietary ecosystems and licensing limitations, rendering them impractical for academic research.
Furthermore, prior studies focus on specific applications leveraging energy harvesting technologies, including novel wireless networks and systems~\cite{wisp,naderiparizi2015wispcam, shen202423,kuo202321,Flync,feng2025deciphering, iyer2022wind}, environmental monitoring and sensing~\cite{everactive,millimobile, saffari2022smart,ewssz,he2024hornbill,fonseca2023system,FarmBeats,weiguo2025sigcomm, dsouza2024densor}, health monitoring~\cite{li2024cyclops, li2022smartlens}, battery-free cameras~\cite{naderiparizi2015wispcam, giordano2020battery, naderiparizi2016battery}, spectrum mapping~\cite{ahmed2023battery}, battery-free testbeds~\cite{riotee, bakar2022protean,hester2017flicker}, and others.
Distinctively, \sysname introduces a paradigm shift in RF front-end design through its innovative ``PLL + RF switch'' architecture, which seamlessly integrates active and passive communication modes. This approach is complemented by a dedicated energy management plane that dynamically regulates power input and output.

\subsection{Hybrid Active-Passive Communication.}
Prior studies~\cite{li2020hybrid,mostafa2022transmit,gu2023computation,li2019adaptive,kan2023adaptive, li2020bandit,long2021achieving,du2021comparing, ma2022covert,chen2019joint,yang2019max, ye2022resource,shi2023wireless,yang2021energy,zargari2022energy, mao2022intelligent} have explored the performance of hybrid active-passive communication through modeling and theoretical approaches, though such efforts have been limited by the absence of a platform for rapidly validating proposed ideas. Although Braidio~\cite{hu2016braidio} introduces an asymmetric energy-aware hardware system, it incorporates dedicated BLE components, which may restrict its adaptability to other use cases.
To the best of our knowledge, \sysname is the first configurable radio platform that natively supports hybrid active-passive communication, offering a versatile foundation for broad research applications.

\section{Conclusion}
%

This paper introduces \sysname, the first configurable radio platform purpose-built to address the fragmented landscape of A-IoT. By proposing a novel parameterized architecture, \sysname successfully unifies active and passive communication modes within a single RF front-end, sharing a generic baseband to minimize hardware redundancy. Furthermore, the platform integrates a scalable energy management plane capable of orchestrating multi-source energy harvesting for battery-free operation. Extensive evaluations across 3GPP A-IoT, IEEE 802.11AMP, and Bluetooth SIG protocols demonstrate that \sysname achieves performance parity with dedicated hardware. As a fully open-source testbed, \sysname serves as a critical enabler for the community to validate emerging protocols and accelerate the standardization of next-generation A-IoT networks.

\bibliographystyle{IEEEtran}
\bibliography{reference}

\end{document}